\documentclass[
superscriptaddress,
groupedaddress,
preprintnumbers,
nofootinbib,
amsmath,amssymb,
aps,
prd,
floatfix,
]{revtex4-2}
\usepackage[utf8]{inputenc}
\usepackage{graphicx}
\usepackage{dcolumn}
\usepackage{bm}
\usepackage{hyperref}
\hypersetup{
    colorlinks=true,
    allcolors=blue,
}

\newcommand{\eq}[1]{Eq.~\eqref{#1}}

\usepackage{xcolor}

\begin{document}

\title{New Exact Vacuum Solutions in Extended Bumblebee Gravity}
\author{Jie Zhu}
 \email{jiezhu@cqu.edu.cn}
 \affiliation{Department of Physics and Chongqing Key Laboratory for Strongly Coupled Physics, Chongqing University, Chongqing 401331, P.R. China}

\author{Hao Li}
  \email{Corresponding author: haolee@cqu.edu.cn}
   \affiliation{Department of Physics and Chongqing Key Laboratory for Strongly Coupled Physics, Chongqing University, Chongqing 401331, P.R. China}

\date{\today}

\begin{abstract}
We investigate the static spherically symmetric vacuum solutions in a generalized bumblebee gravity model characterized by non-minimal couplings $B^2 R$ and $B^\mu B^\nu R_{\mu\nu}$. We demonstrate that the variation of the action and the imposition of the vacuum expectation value constraint are non-commutative, leading to a richer solution space than previously explored. 
A diverse set of solutions, including naked singularities, black holes, and wormholes, is obtained, and as many as ten exact solutions are presented.
The thermodynamic properties of the new black hole solutions are also analyzed, and a subset of these solutions is found to have zero entropy.
We argue that if such a non-minimally coupled vector-tensor gravity provides a fundamental description of the universe, it is best described by a Bumblebee-type theory, where the vector field acquires a VEV.
\end{abstract}

\maketitle

\section{Introduction}

The General Relativity (GR) of Einstein has achieved remarkable success in describing gravitational phenomena across a vast range of scales, from Solar System tests to the recent detections of gravitational waves by LIGO/Virgo. Despite its robustness, GR faces significant challenges, particularly when integrated with the Standard Model of particle physics, leading to the pursuit of modified gravity theories. Among various extensions, Lorentz invariance violation (LV) has emerged as a promising window into the potential quantum nature of gravity at the Planck scale.

Originating from string theory~\cite{Kostelecky:1988zi}, the notion of spontaneous Lorentz symmetry breaking laid the theoretical foundation for the Standard-Model Extension (SME), a comprehensive toolkit designed to characterize Lorentz-violating effects~\cite{Colladay:1996iz, Colladay:1998fq, Colladay:2001wk, Kostelecky:2000mm, Kostelecky:2003fs}.
A well-motivated framework for studying LV in the gravity sector is the bumblebee model~\cite{Kostelecky:2003fs, Bumblebee05, Bumblebee08}, where a vector field $B_\mu$ acquires a non-zero vacuum expectation value (VEV) through a spontaneous symmetry breaking (SSB) mechanism triggered by a potential. The VEV of the bumblebee field defines a preferred direction in spacetime, thereby naturally incorporating LV into the gravitational action. Early explorations of this model primarily focused on minimal couplings or a specific non-minimal coupling term of the form $\frac{\xi}{2\kappa} B^\mu B^\nu R_{\mu\nu}$. 
In 2017, Casana et al. reported the first exact Schwarzschild-like black hole solution in bumblebee gravity~\cite{Casana:2017jkc}, and the classification of the static spherical solutions of the bumblebee gravity has been reported recently~\cite{Li:2025rjv, Zhu:2025fiy, Liu:2025oho}.
A wide variety of further generalizations of solutions can also be found in Refs.~\cite{Ovgun:2018xys, Maluf:2020kgf, Ding:2019mal, Santos:2014nxm, Jha:2020pvk, Filho:2022yrk, Xu:2022frb, Ding:2023niy, Liu:2024axg, Bailey:2025oun, Li:2025tcd}.

However, a more generalized vector-tensor coupling scheme, including terms such as $\frac{\lambda}{2\kappa} B^2 R$, remains relatively under-explored. A critical conceptual subtlety often overlooked in the literature is the non-commutativity between the variational procedure and the imposition of the VEV constraint. In models where the vector field is strictly constrained by a potential, the order in which one varies the action and fixes the field magnitude can lead to fundamentally different field equations. This non-commutative nature suggests that the solution space of extended bumblebee gravity is far richer than previously recognized, potentially harboring novel gravitational objects that have escaped detection in simpler models.

In this work, we provide a comprehensive classification of static spherically symmetric vacuum solutions within this extended bumblebee framework. By rigorously accounting for the non-minimal couplings $\lambda$ and $\xi$, we derive nine distinct types of exact solutions, ranging from black holes and naked singularities to exotic traversable wormholes. To further investigate the physical consistency of these solutions, we perform a thermodynamic analysis using the Iyer-Wald formalism. Our results reveal a notable discrepancy between the standard Wald entropy formula and the Iyer-Wald results in certain sectors of the theory. Finally, we contrast the ``bumblebee" model with its ``free" vector-tensor counterpart, demonstrating that the presence of a potential is not merely a formal requirement but a necessary ingredient for a physically predictive theory.
As a demonstration, we further develop a highly non-trivial traversable wormhole solution for the $\xi=0$ free theory.

The paper is organized as follows.
Section~\ref{sec:Bumblebee} reviews the framework of the extended bumblebee gravity.
In Sec .~\ref{sec:solve}, we show the details of the solving procedure.
Section~\ref{sec:solutions} summarizes the solutions.
In Sec.~\ref{sec:thermodynamics}, we perform the Iyer–Wald formalism to show the thermodynamics of the newly obtained black hole solutions.
Finally, Section~\ref{sec:discusion} discusses and summarizes our results.

\section{Extended bumblebee gravity}\label{sec:Bumblebee}
Here, we consider the action of the general bumblebee gravity as 
\begin{equation}
S=\int d^4x\sqrt{-g}\left(\frac{1}{2\kappa}\left(R+\lambda B_\mu B^\mu R + \xi B^\mu B^\nu R_{\mu\nu}\right)-\frac{1}{4}B_{\mu\nu}B^{\mu\nu}-V\right)+S_\mathrm{m},  \label{eq:action}
\end{equation}
where $g$ is the determinant of the metric $g_{\mu\nu}$, the constant $\kappa\equiv8\pi G$ with $G$ being the gravitational constant, $S_{\mathrm{m}}$ represents the action for matter fields of no interest in this work, $B_\mu$ is the bumblebee field, and the field strength tensor is $B_{\mu\nu}=\partial_{\mu}B_{\nu}-\partial_{\nu}B_{\mu}$.
In bumblebee theories, the potential $V$ is selected to provide a non-vanishing VEV for $B_\mu$, and could have the following general functional form
\begin{equation*}
    V\equiv V(B^\mu B_\mu + s b^2),\label{potential}
\end{equation*}
where $b$ is a positive real constant, and $s=\pm 1$ or 0 to determine whether the expection value of $B_\mu$ is timelike, spacelike or lightlike. 
In the literature, it is usually assumed that \(V\) has (at least one of) its minimum/maximum at \(0\), thus
\begin{equation*}
    V(0)=0,\ \text{and}\ V'(0)=0.\label{vacuumcondition}
\end{equation*}
The VEV of the bumblebee field is determined when $V(B^\mu B_\mu + s b^2)=0$, implying that
\begin{equation*}
    B^\mu B_\mu+s b^2=0, 
\end{equation*}
The above equation provides a non-null vacuum expectation value
\begin{equation*}
    \langle B^\mu\rangle=b^\mu,
\end{equation*}
where $b_\mu b^\mu + s b^2=0$.
The equation of motions for $g_{\mu\nu}$ is $G_{\mu\nu}=R_{\mu\nu}-\frac{1}{2}Rg_{\mu\nu}=\kappa T_{\mu\nu}$, or 
\begin{equation}
R_{\mu\nu}-\kappa(T_{\mu\nu}-\frac{1}{2}g_{\mu\nu}T)=0, \label{eq:EOMg}
\end{equation}
where 
\begin{equation*}
T_{\mu\nu}=T_{\mu\nu}^{\mathrm{M}}+T_{\mu\nu}^{B},
\end{equation*}
$T_{\mu\nu}^{\mathrm{M}}$ is the energy-momentum tensor of matter, and 
\begin{equation}
\begin{aligned}
T_{\mu\nu}^{B}=&
-B_{\mu\alpha}B_{\nu}^{\alpha}
-\frac{1}{4}B_{\alpha\beta}B^{\alpha\beta}g_{\mu\nu}
-Vg_{\mu\nu}+2V^{\prime}B_{\mu}B_{\nu}
+\frac{\xi}{\kappa}\left(
\frac{1}{2}B^{\alpha}B^{\beta}R_{\alpha\beta}g_{\mu\nu}
-B_{\mu}B^{\alpha}R_{\alpha\nu}
-B_{\nu}B^{\alpha}R_{\alpha\mu}\right.
\\&
+\frac{1}{2}\nabla_{\alpha}\nabla_{\mu}\left(B^{\alpha}B_{\nu}\right)
+\frac{1}{2}\nabla_{\alpha}\nabla_{\nu}\left(B^{\alpha}B_{\mu}\right)
\left.
-\frac{1}{2}\nabla^{2}\left(B_{\mu}B_{\nu}\right)
-\frac{1}{2}g_{\mu\nu}\nabla_{\alpha}\nabla_{\beta}\left(B^{\alpha}B^{\beta}\right)\right)
\\&
+\frac{\lambda}{\kappa}\left(
-B_\alpha B^\alpha G_{\mu\nu}
-B_\mu B_\nu R
\right.
\left.
+2\nabla_\nu(B_\alpha \nabla_\mu B^\alpha)
-2g_{\mu\nu} \nabla_\alpha(B_\beta \nabla^\alpha B^\beta)\right).
\end{aligned}
\end{equation}
For the $B_\mu$ sector, the equation of motion is
\begin{equation}
\nabla^{\mu}B_{\mu\nu}
+\frac{1}{\kappa}(\xi B^\mu R_{\mu \nu}+\lambda B_\nu R)-2V' B_\nu =0.\label{eq:EOMB}
\end{equation}

\section{Solving the equations}\label{sec:solve}

In this section, we solve the static spherical vacuum solutions to the vector-tensor model \eq{eq:action}.
 We are interested in the vacuum solutions, namely $T_{\mu\nu}^{\mathrm{M}}$, in the sense that no other kind of matter exists, besides the bumblebee field, which might be composed of some type of unknown matter. 
Also, we consider the case that $B_\mu$ admits VEV as $b_\mu$ and there is no cosmological constant, so for $B_\mu$ and $V$ we have
\begin{equation*}
    B_\mu B^\mu+sb^2=0,\quad V(0)=0, \quad\text{and}~ V'(0)=0.
\end{equation*}
It is worth noting that, in this case, the equations of motion are identical to those where $V \equiv 0$. Consequently, the solutions can be interpreted either as general solutions to the general bumblebee model where vacuum expectation values are derived from the potential $V(B^\mu B_\mu + s b^2)$, or as specific solutions to the vector-tensor theory with a vanishing potential.

One might question why, similar to Einstein-Aether theory, one cannot impose the condition $B_\mu B^\mu + sb^2 = 0$ to transform $B_\mu B^\mu R$ into $R$ within the action, which would effectively manifest as a redefinition of the parameter $\kappa$ and $\xi$ prior to variation. 
In fact, in Einstein-Aether theory, the constraint $u^\mu u_\mu = \pm 1$ is derived by varying the action with respect to a Lagrange multiplier; thus, the order of substituting the constraint and performing the variation is interchangeable. 
However, in the present case, $B_\mu B^\mu + sb^2 = 0$ arises from the potential $V$ reaching its VEV. Under these circumstances, the operations of variation and imposing the constraint do not commute.
Furthermore, we shall demonstrate below the existence of new vacuum solutions that extend beyond those previously discovered; these novel solutions precisely illustrate that the two operations are non-commutative.

Before proceeding with the derivation, we first provide a preliminary classification of the vacuum solutions.
Similar to the discussion in Refs~\cite{Ji:2024aeg, Li:2025rjv}, the static spherical vacuum solution to Bumblebee Gravity can be classified into two classes. For the static spherical field configuration \(B_\mu=b_\mu=(b_t(r), b_r(r), 0, 0)\), the only non-vanishing components of field strength tensor $b_{\mu\nu}=\partial_\mu b_\nu-\partial_\nu b_\mu$ are \(b_{rt}=-b_{tr}=b_t'(r).\)
So \eq{eq:EOMB} is
\begin{equation}
\frac{1}{\sqrt{-g}}\partial_{r}(\sqrt{-g}b^{\bar{\mu}r})
-\frac{1}{\kappa}b_{\bar{\mu}}(\xi R^{\bar{\mu}}_{\bar{\mu}}+\lambda R)=0,
\end{equation}
where there is no summation for the index \(\bar{\mu}=t,r,\theta,\phi\),
we can easily see that the $r-$~component of the above equation is
\begin{equation}
b_{r}(\xi R^{r}_r+\lambda R)=0.\label{eq:Rrrbr}
\end{equation}
\eq{eq:Rrrbr} shows that the spherically symmetric vacuum solutions in the bumblebee model can be classified into two \emph{disjoint} classes:
\begin{itemize}
    \item Class I: $b_r \equiv 0$,
    \item Class II: $\xi R^{r}_r+\lambda R \equiv 0$.
\end{itemize}
For the space-like VEVs, $b_r$ cannot be identically zero, so the solutions are classified as belonging to Class II,
whereas in the time-like cases, $b_r\equiv 0$ can be achieved.
From the expression of the Class II condition, it is already evident that we cannot impose the condition $B_\mu B^\mu + sb^2 = 0$ to transform $B_\mu B^\mu R$ into $R$ within the action.

The general static spherical metric can be expressed as 
\begin{equation}
ds^2=-G(r) dr^2+\frac{1}{H(r)}dr^2+ R(r)^2 
d\Omega^2, \label{eq:metdef}
\end{equation}
and the general static spherical background $b_\mu$ can be expressed as
\begin{equation}
    b_\mu=(b_t(r),b_r(r),0,0). \label{eq:b_s}
\end{equation}
If $B_\mu=b_\mu$ is a solution, then $B_\mu=-b_\mu$ is also a solution.
In this work, we only consider one branch.
One retains the coordinate gauge freedom to further simplify the metric components, such as $R(r) \equiv r$ or $G(r) = H(r)$.
In the subsequent calculations, we will switch between these two coordinate choices as needed; however, for the time being, we shall retain this gauge freedom. 
In the following, let $\mathcal{E}_{\mu\nu}$ and $\mathcal{E}_{\mu}$ denote the LHS of the components of the metric equations of motion \eq{eq:EOMg} and the $B_\mu$ equations of motion \eq{eq:EOMB}, respectively. 
Also, for simplicity, we define two parameters $\ell_1\equiv \xi b^2$ and $\ell_2\equiv \lambda b^2$.
From the action \eq{eq:action} and the condition $B^\mu B_\mu+sb^2=0$, it seems that $\ell_1$ is related to the strength of Lorentz symmetry breaking, whereas $\ell_2$ is associated with the effective gravitational constant. 
Since the case of $\lambda = 0$ ($\ell_2= 0$) has been thoroughly explored in previous literature~\cite{Casana:2017jkc, Li:2025rjv, Zhu:2025fiy, Liu:2025oho}, in this work, we only consider the case that $\ell_2\neq 0$.

\subsection{Class I Solutions}

For the Class I solutions, $B_r=0$ and the VEV of $B_\mu$ should be timelike, so we have $s=1$.
In this case, in the analysis of the equation of motion, we use the coordinate so that $R(r)=r$.
We also use the new function $\alpha(r)$ and $\beta(r)$ so that $G(r)=e^{2\alpha(r)}$ and $H(r)=e^{-2\beta(r)}$, and the metric becomes $ds^2=-e^{2\alpha(r)} dr^2+e^{2\beta(r)}dr^2+ r^2 d\Omega^2$.
To satisfy the constraint $B_\mu B^\mu=-b^2$, we have $B_t = b e^{\alpha(r)}$.
Under the field configurations listed above, the non-vanishing components of \eq{eq:EOMg} and \eq{eq:EOMB} are
\begin{align}
\mathcal{E}_{tt}=&
-r^2 \left(\kappa b^2  +\ell_1+4 \ell_2-2\right) \alpha '^2
-\left(\ell_1+4 \ell_2-2\right) r^2 \alpha ''
\nonumber\\&
+\alpha ' \left(\left(\ell_1+4 \ell_2-2\right) r^2 \beta '
-2 \left(\ell_1+4 \ell_2-2\right) r\right)
+4 \ell_2 r \beta '+2 \ell_2 \left(e^{2 \beta }-1\right),\label{eq:tt1}
\\
\mathcal{E}_{rr}=&
r^2 \left(-\kappa b^2+\ell_1+2\right) \alpha '^2+\left(\ell_1+2\right) r^2 \alpha ''+\alpha ' \left(2 \left(\ell_1+2 \ell_2\right) r-\left(\ell_1+2\right) r^2 \beta '\right)-2 \ell_2 \left(e^{2 \beta }-1\right)-4 r \beta ',\label{eq:rr1}
\\
\mathcal{E}_{\theta\theta}=&
-r^2 \left(\kappa b^2  +\ell_1+2 \ell_2\right) \alpha '^2
-\left(\ell_1+2 \ell_2\right) r^2 \alpha ''+\alpha ' \left(\left(\ell_1+2 \ell_2\right) r^2 \beta '-2 \left(\ell_1+\ell_2+1\right) r\right)
\nonumber\\&
+2 \left(\ell_2+1\right) r \beta '+2 \left(e^{2 \beta }-1\right),\label{eq:thth1}
\\
\mathcal{E}_{\phi\phi}=& \sin^2(\theta) \mathcal{E}_{\theta\theta},
\\
\mathcal{E}_t=&
r^2 \left(b^2 \kappa -\ell_1-2 \ell_2\right) \alpha ''
+\alpha ' \left(r^2 \left(b^2 (-\kappa )+\ell_1+2 \ell_2\right) \beta '+2 r \left(b^2 \kappa -\ell_1-2 \ell_2\right)\right)
\nonumber\\&
-\left(\left(\ell_1+2 \ell_2\right) r^2 \alpha '^2\right)+4 \ell_2 r \beta '+2 \ell_2 \left(e^{2 \beta }-1\right).\label{eq:t1}
\end{align}
This constitutes an overdetermined system, since we have only two variables, $\alpha(r)$ and $\beta(r)$, yet there are four independent equations.
The above equations can be simplified by the following steps:
\begin{enumerate}
\item By taking linear combinations of these four equations, we eliminate the term $e^{2\beta}$, resulting in three independent equations;
\item By taking linear combinations of the three equations from the last step, we eliminate the term $\alpha'(r)\beta'(r)$, resulting in two independent equations;
\item By taking linear combinations of the remaining two equations, we eliminate the term $\beta'(r)$, resulting in one equation.
\end{enumerate}
After the three steps listed above, we obtain a simple relation
\begin{equation*}
4 b^2 \kappa  \left(\ell_2-1\right){}^2 \left(\kappa b^2  +2 \ell_2-2\right) r^3 \alpha '^2=0.
\end{equation*}
The solution in this case depends on which part of the relation is zero.

\subsubsection{\texorpdfstring{Case of $\alpha'(r)=0$}{Case of alpha'(r)=0}}

If $\ell_2\neq1$ and $\kappa b^2  +2 \ell_2-2\neq 0$, we have $\alpha '(r)=0$, so $\alpha(r)$ is a constant, which implies that $g_{tt}=e^{2\alpha(r)}$ is also a constant. 
By the redefinition of $t$, we can choose $\alpha(r)=0$.
Substitute $\alpha(r)=0$ into Eqs.~(\ref{eq:tt1}),~(\ref{eq:rr1}),~(\ref{eq:thth1}), and (\ref{eq:t1}), we have the following equations
\begin{align*}
2 r \beta '+e^{2 \beta }-1=0,\\
\left(\ell_2+1\right) r \beta '+e^{2 \beta }-1=0,\\
2 r \beta '+\ell_2 \left(e^{2 \beta }-1\right)=0.
\end{align*}
To obtain a non-trivial solution ($\beta(r)\neq 0$), we need $\ell_2\equiv \lambda b^2=1$, so $b= 1/\sqrt{\lambda}$.
Now the solution for $\beta(r)$ is
\begin{equation*}
\beta(r)=-\frac{1}{2}\ln\left(1-\frac{R_s}{r}\right ),
\end{equation*}
where $R_s$ is a constant.
So the first solution is
\begin{equation}
ds^2=-dt^2+\left(1-\frac{R_s}{r}\right )^{-1} dr^2+r^2d\Omega^2, \quad b_t = \frac{1}{\sqrt{\lambda}}, \quad b_r = 0.
\end{equation}
It is easy to check that this solution satisfies all of the equations.
In this solution, the timelike VEV is $b = \frac{1}{\sqrt{\lambda}}$, and this solution exists only in the regime where $\lambda > 0$.
Although this solution requires a fixed value for $b$, as previously discussed, it can be regarded as a particular solution to the vector-tensor theory with $V\equiv0$.

\subsubsection{\texorpdfstring{Case of $\kappa b^2  +2 \ell_2-2= 0$}{Case of kappa b\^2+2 l\_2-2=0}}

Here, we consider the case  $\kappa b^2  +2 \ell_2-2= 0$. 
Using the definition of $\ell_1$ and $\ell_2$, in this case, we have
\begin{equation*}
\ell_1=\frac{2\xi}{\kappa+2\lambda},\quad 
\ell_2 = \frac{2\lambda}{\kappa+2\lambda},\quad\text{and}~
b=\sqrt{\frac{2}{\kappa+2\lambda}}=\sqrt{\frac{2(1-\ell_2)}{\kappa}}.
\end{equation*}
Substitute $b=\sqrt{\frac{2(1-\ell_2)}{\kappa}}$ into Eqs.~(\ref{eq:tt1}),~(\ref{eq:rr1}),~(\ref{eq:thth1}), and (\ref{eq:t1}), and take linear combinations of these four equations to eliminate the term $e^{2\beta}$, we obtain only two independent equations as
\begin{equation}
-r \alpha ''+\alpha ' \left(r \beta '-1\right)+\beta '=0,\label{eq:ab1}
\end{equation}
\begin{equation}
-\left(\ell_1+2 \ell_2\right) r \alpha ''+\alpha ' \left(\left(\ell_1+2 \ell_2\right) r \beta '-2 \left(\ell_1+\ell_2-1\right)\right)+\ell_1 (-r) \alpha '^2+2 \left(\ell_2+1\right) \beta '=0.\label{eq:ab2}
\end{equation}
The interesting part is that, by applying $2\ell_2\times$\eq{eq:ab1}$-$\eq{eq:ab2}, we have
\begin{equation}
\ell_1 r \alpha ''+\alpha ' \left(2 \left(\ell_1-1\right)-\ell_1 r \beta '\right)+\ell_1 r \alpha '^2-2 \beta '=0,\label{eq:ab3}
\end{equation}
which depend on $\ell_1$ only.
As a consequence, any solution in this case should also be the solution of the special case $\ell_2=0$, or $\lambda=0$, with differences only in the value of $b$.
Luckily, in Ref.~\cite{Li:2025rjv} we already have the solutions for the case $\lambda=0$, so we can directly use the results and write down the solutions.
The first solution in this case is
\begin{gather*}
ds^2=-\left(\frac{r}{R_s}\right)^{\frac{2(2-\ell_1)}{\ell_1}} dt^2+\frac{4}{\ell_1^2}dr^2+r^2 d\Omega^2,
\\
\quad b_t = b \left(\frac{r}{R_s}\right)^{2/\ell_1-1},
\quad b_r = 0,
\end{gather*}
and the second solution is 
\begin{gather*}
ds^2= -\left(1-\frac{R_s}{r}\right)^{2-\ell_1} dt^2 
+\left(1-\frac{R_s}{r}\right)^{-2+\ell_1} dr^2
+\left(1-\frac{R_s}{r}\right)^{\ell_1}r^2d\Omega^2,
\\
b_t =b \left(1-\frac{R_s}{r}\right)^{1-\ell_1/2},
\quad b_r=0,
\end{gather*}
where
\begin{equation}
\ell_1=\frac{2\xi}{\kappa+2\lambda}, \quad \text{and}~ b=\sqrt{\frac{2}{\kappa+2\lambda}}.
\end{equation}
One can check that these two equations satisfy the equations of motion~(\ref{eq:EOMg}) and (\ref{eq:EOMB}).
In this case, these two solutions can be interpreted as the result of substituting the constraint $B_\mu B^\mu = -b^2$ into the action and subsequently redefining the parameter $\kappa$ by
\begin{equation*}
\frac{1}{2\kappa'} R = \frac{1}{2\kappa}(R+\lambda B_\mu B^\mu R)=\frac{1-\ell_2}{2\kappa} R,
\end{equation*}
and replace $\kappa\to\kappa'=\kappa/(1-\ell_2)$ in the solution obtained in Ref.~\cite{Li:2025rjv}.

\subsection{Class II Solutions}

In this section, we consider the Class II solutions.
In this case, by a suitable choice of coordinates, we set $H(r) = G(r)$.
To satisfy the constraint $B_\mu B^\mu=-sb^2$, we use the field configuration as 
\begin{equation}
b_t = b \sqrt{G(r)} U(r), \quad b_r = b\frac{\sqrt{U(r)^2-s}}{\sqrt{G(r)}},
\end{equation}
where $U(r)$ is a function to be determined.
Since $b_r\neq 0$, so $U(r)^2-s\neq 0$.
Under this field configuration, the non-vanishing component of the equation of motion for the $B_\mu$ sector is
\begin{align}
\mathcal{E}_t=&
-2\left(-\kappa b^2 +\ell_1+2 \ell_2\right) G R^2 U  G''
+G' \left(4 b^2 \kappa  G R^2 U'-4 G R U \left(-\kappa b^2 +\ell_1+4 \ell_2\right) R'\right)
\nonumber\\&
-b^2 \kappa  R^2 U G'^2
+8 b^2 \kappa  G^2 R R' U'
+4 b^2 \kappa  G^2 R^2 U''
-16 \ell_2 G^2 R U R''
-8 \ell_2 G^2 U R'^2+8 \ell_2 G U,\label{eq:t2}
\\
\mathcal{E}_r=&
\sqrt{U^2-s}\left(\left(\ell_1+2 \ell_2\right) R^2 G''+2 \left(\ell_1+4 \ell_2\right) R G' R'+4 \left(\ell_1+2 \ell_2\right) G R R''+4 \ell_2 G R'^2-4 \ell_2\right).
\end{align}
For the metric sector, the equations are
\begin{align}
\mathcal{E}_{tt}=&
G' \left(-4 G R^2 U \left(\kappa b^2 -4 \ell_1\right) U'-4 G R R' \left(2 \ell_2 \left(4 U^2-s\right)+\ell_1 s-2\right)\right)
-b^2 \kappa  R^2 U^2 G'^2
-4 G^2 R^2 \left(b^2 \kappa -2 \ell_1\right) U'^2
\nonumber\\&
-2 G R^2 G'' \left(4 \ell_2 U^2+\ell_1 s-2\right)
+8 G^2 R R'' \left(2 \ell_2 \left(s-2 U^2\right)+\ell_1 \left(s-U^2\right)\right)
+8 \ell_2 G^2 R'^2 \left(s-2 U^2\right)
\nonumber\\&
+16 \ell_1 G^2 R U R' U'
+8 \ell_1 G^2 R^2 U U''
-8 \ell_2 G \left(s-2 U^2\right),
\\
\mathcal{E}_{rr}=&
G' \left(4 G R^2 U \left(\kappa b^2 -4 \ell_1\right) U'+4 G R R' \left(-4 \left(\ell_1+2 \ell_2\right) U^2+3 \ell_1 s+6 \ell_2 s-2\right)\right)
+b^2 \kappa  R^2 U^2 G'^2
\nonumber\\&
+4 G^2 R^2 \left(b^2 \kappa -2 \ell_1\right) U'^2
+2 G R^2 G'' \left(-4 \left(\ell_1+\ell_2\right) U^2+3 \ell_1 s+4 \ell_2 s-2\right)
+8 G^2 R R'' \left(\left(3 \ell_1+4 \ell_2\right) \left(s-U^2\right)-2\right)
\nonumber\\&
+8 \ell_2 G^2 R'^2 \left(s-2 U^2\right)
-16 \ell_1 G^2 R U R' U'
-8 \ell_1 G^2 R^2 U U''
-8 \ell_2 G \left(s-2 U^2\right),
\\
\mathcal{E}_{\theta\theta}=&
G' \left(4 G R R' \left(\ell_1 \left(s-2 U^2\right)-2 \left(\ell_2 s+1\right)\right)-4 b^2 \kappa  G R^2 U U'\right)
-b^2 \kappa  R^2 U^2 G'^2
-4 b^2 \kappa  G^2 R^2 U'^2
\nonumber\\&
-2 \left(\ell_1+2 \ell_2\right) s G R^2 G''
-8 G^2 \left(\ell_2 s+1\right) R R''
+8 G^2 R'^2 \left(\ell_1 \left(s-U^2\right)-1\right)
-16 \ell_1 G^2 R U R' U'+8 G,
\\
\mathcal{E}_{\phi\phi}=&\sin^2(\theta)\mathcal{E}_{\theta\theta},
\\
\mathcal{E}_{tr}=&U(r)\mathcal{E}_{r}.
\end{align}
Now by applying $\mathcal{E}_{tt}+\mathcal{E}_{rr}-4 G (s- 2U^2)/{\sqrt{U^2-s}} \mathcal{E}_r=0$, we have
\begin{equation}
16 G^2 \left(\left(\ell_1+\ell_2\right) s-1\right) R R''=0.\label{eq:Req}
\end{equation}
So if $(\ell_1+\ell_2)s\neq1$, we have $R''(r)=0$. 
Without losing generality, the solution is $R(r)=r$, since we can always redefine $t$ and $r$ such that the metric retains the form with $G(r)=H(r)$.
With $R(r)=r$, the equation $\mathcal{E}_r=0$ becomes
\begin{equation}
\left(\ell_1+2 \ell_2\right) r^2 G''(r)+2 \left(\ell_1+4 \ell_2\right) r G'(r)+4 \ell_2 G(r)=4\ell_2. \label{eq:G}
\end{equation}
This is a differential equation of the Euler–Cauchy type,
and the solution for is
\begin{equation}
G(r)=1-\frac{R_s}{r}+\frac{Q}{r^p}, \quad  p=\frac{4\ell_2}{\ell_1+2\ell_2}=\frac{4\lambda}{\xi+2\lambda},
\end{equation}
where $R_s$ and $Q$ are free parameters.
If $\lambda=0$, then $p=0$, and we have $G(r)= 1+Q-\frac{R_s}{r}=(1+Q)(1-\frac{R_s'}{r})$, and the metric reduces to the form found in Refs.~\cite{Casana:2017jkc, Zhu:2025fiy, Liu:2025oho}.
However, for $\lambda \neq 0$, the resulting metric solution differs fundamentally from the $\lambda=0$ case. This discrepancy highlights the fact that the constraint $B_\mu B^\mu = -sb^2$ cannot be substituted back into the action prior to performing the variation.

Substitute $R(r)=r$ and the solution for $G(r)$ in the equation $\mathcal{E}_t=0$, we can obtain the equation for $U(r)$, and the solution is
\begin{equation*}
U(r)=\frac{c_1+\frac{c_2}{r}}{\sqrt{1-\frac{R_s}{r}+\frac{Q^p}{r^p}}}.
\end{equation*}
So we have
\begin{equation*}
b_t(r)=b \sqrt{G(r)} U(r)=b\left(c_1 +\frac{c_2}{r}\right).
\end{equation*}
We need to check other equations. 
Substitute the solution for $R(r)$, $G(r)$ and $U(r)$ into $\mathcal{E}_{tt}$, $\mathcal{E}_{rr}$, and $\mathcal{E}_{\theta\theta}$, we have
\begin{gather*}
\mathcal{E}_{tt}=-\mathcal{E}_{rr}\propto
\frac{c_2{}^2 \left(2 \ell_1-b^2 \kappa \right)}{r^2}
+\frac{2Q \left(\ell_1-2 \ell_2\right)  \left(\left(\ell_1+\ell_2\right) s-1\right)}{(\ell_1+2 \ell_2)r^p},
\\
\mathcal{E}_{\theta\theta}\propto
\frac{c_2{}^2 \left(2 \ell_1-b^2 \kappa \right)}{r^2}
+\frac{2Q \left(\ell_1-2 \ell_2\right)  \left(\left(\ell_1+\ell_2\right) s-1\right)}{(\ell_1+2 \ell_2)r^p}
+2 \ell_1 \left(s-c_1{}^2\right).
\end{gather*}
The constraints on the parameters depend on the value $p=4\lambda/(\xi+2\lambda)$.

\subsubsection{\texorpdfstring{Case of $\xi =0$}{Case of xi = 0}}

We first consider the case $p=2$, or $\xi=\ell_1=0$.
In this case, we first replace $Q$ with $Q^2$, and the requirement that $\mathcal{E}_{tt}=\mathcal{E}_{rr}=\mathcal{E}_{\theta\theta}=0$ becomes
\begin{equation*}
\kappa b^2 c_2^2+2Q^2 (s\ell_2-1)=0,
\end{equation*}
and we have 
\begin{equation*}
c_2 = \frac{Q}{b}\sqrt{\frac{2(1-s\ell_2)}{\kappa}}.
\end{equation*}
In this case, the solution is
\begin{gather*}
ds^2=-f dt^2+\frac{dr^2}{f}+r^2 d\Omega^2, \quad f= 1-\frac{R_s}{r}+\frac{Q^2}{r^2},
\\
b_t= P + \sqrt{\frac{2(1-\lambda s b^2)}{\kappa}}\frac{Q}{r},
\quad b_r  =\frac{\sqrt{b_t^2-sb^2 f}}{f},
\end{gather*}
where $P$, $Q$, $R_s$ and $b$ are free parameters.
In fact, this solution can be viewed as the result of performing a gauge transformation on the vector field of the Reissner–Nordstr{\"o}m (RN) solution, followed by a redefinition of the parameter $\kappa$ as $\kappa'=\frac{\kappa}{1-\lambda s b^2}=\frac{\kappa}{1-s\ell_2}$.

\subsubsection{\texorpdfstring{Case of $\lambda=-\xi/2$}{Case of lambda=- xi/2}}\label{sec:sol1}

In this case, $\lambda=-\xi/2$, and the coupling in the action becomes
\begin{equation*}
\frac{1}{2\kappa}\left(-\frac{\xi}{2}B_\mu B^\mu R+\xi B_\mu B_\nu R^{\mu\nu}\right)=\frac{\xi}{2\kappa}G_{\mu\nu}B^\mu B^\nu.
\end{equation*}
That is, the coupling between $B_\mu$ and gravity in the action is directly manifested as the coupling between $B_\mu$ and the Einstein tensor.
In this case, \eq{eq:G} becomes
\begin{equation*}
r G'(r)+G(r)-1=0,
\end{equation*}
and the solution is
\begin{equation*}
G(r)=1-\frac{R_s}{r}.
\end{equation*}
So in this case, the metric solution reduces exactly to the Schwarzschild solution.
With $R(r)=r$, $\ell_2=-\ell_1/2$ and the solution of $G(r)$ substituting in \eq{eq:t2}, we obtain the equation for $U(r)$ as
\begin{equation*}
4 r^2 (r-R_s)^2 U''(r)+4 r (2 r-R_s) (r-R_s) U'(r)-R_s^2 U(r)=0,
\end{equation*}
and the solution is
\begin{equation*}
U(r) = \frac{c_1 r+c_2}{\sqrt{r(r-R_s)}}.
\end{equation*}
Now we have $b_t(r)=b \sqrt{G(r)} U(r)=b(c_1 +\frac{c_2}{r})$, 
so if the solution for $B_\mu$ is physically meaningful, we need $c_1$ and $c_2$ are real number.
Substitute the solutions of $R(r)$, $G(r)$, and $U(r)$, and the relation $\ell_2=-\ell_1/2$ into other equations of motion, we obtain
\begin{align*}
\mathcal{E}_{tt}&=-\mathcal{E}_{rr}=-\frac{4 c_2{}^2 (r-R_s) \left(b^2 \kappa -2 \ell_1\right)}{r^3},
\\
\mathcal{E}_{\theta\theta}&=\frac{4 (r-R_s) \left(c_2{}^2 \left(2 l_1-b^2 \kappa \right)+2 l_1 r^2 \left(s-c_1{}^2\right)\right)}{r^3}.
\end{align*}
So to get a solution, we have the following constraints for the parameter:
\begin{equation*}
c_2{}^2 \left(2 l_1-b^2 \kappa \right)=0,\quad s-c_1{}^2=0.
\end{equation*}
Since $s=\pm 1$ or 0 and $c_1$ is a real number, we have $c_1=s=1$ or 0.
That is, in this case, we only have solutions for \emph{timelike} and \emph{lightlike} VEVs.
If $2\ell_1-\kappa b^2=0$, we have $\xi=\kappa/2$, and the solution is
\begin{gather*}
ds^2= -f dt^2 +\frac{dr^2}{f}+r^2d\Omega^2, \quad  f=1-\frac{R_s}{r},
\\
b_t = b+\frac{Q}{r},\quad b_r =\frac{\sqrt{b_t^2-b^2 f}}{f},
\end{gather*}
where $Q$, $R_s$ and $b$ are free parameters.
If $b=0$, then $B_\mu$ is lightlike. 
In fact, this solution has already been reported in Ref.~\cite{Chagoya:2016aar}.
If $2\ell_1-\kappa b^2\neq 0$, then we have $c_2=0$, and the solution is
\begin{gather*}
ds^2= -f dt^2 +\frac{dr^2}{f}+r^2d\Omega^2, \quad  f=1-\frac{R_s}{r},
\\
b_t = b, \quad b_r =\frac{b\sqrt{R_s r}}{r- R_s},
\end{gather*}
where $R_s$ and $b$ are free parameters.
If $b=0$, then $B_\mu$ is lightlike, but in this case $B_\mu\equiv0$ and the solution is trivial.
Ref.~\cite{Chagoya:2016aar} reported that, in vector-tensor theories with a non-minimal coupling of the form $\frac{\xi}{2\kappa} G_{\mu\nu}B^\mu B^\nu$, the requirement of asymptotic flatness and a non-trivial profile for $b_t$ uniquely fixes the coupling constant to $\xi=\kappa/2$. 
We find that although the configuration $b_t = b$ appears trivial, the resulting spacetime geometry is non-trivial.

\subsubsection{\texorpdfstring{Case of $\lambda=\xi/2$}{Case of lambda= xi/2}}

In the case $\xi\neq 0$, to let $\mathcal{E}_{\theta\theta}$ be identically zero, we have $s=c_1^2$.
Again, since $s=-1, 0, 1$  and $c_1$ is a real number, we have $c_1=s=1$ or $c_1=s=0$.
So in this case, we also only have solutions for \emph{timelike} or \emph{lightlike} VEVs.
We now consider the special case $p=1$, which means that $\ell_1=2\ell_2$ and $\xi=2\lambda$.
In this case, the metric reduces to the Schwarzschild metric, and we can let $Q=0$.
The requirement that $\mathcal{E}_{tt}=\mathcal{E}_{rr}=\mathcal{E}_{\theta\theta}=0$ becomes $c_2{}^2 (2 \ell_1-b^2 \kappa )=0$.
If $2 \ell_1-b^2 \kappa=0$, we have $\xi=\kappa/2$ and $\lambda=\kappa/4$, and the solution is
\begin{gather*}
ds^2=-f dt^2+\frac{dr^2}{f}+r^2 d\Omega^2, \quad f= 1-\frac{R_s}{r},
\\
b_t = b+\frac{Q}{r},\quad b_r =\frac{\sqrt{b_t^2-b^2 f}}{f},
\end{gather*}
where $b$, $Q$ and $R_s$ are free parameters.
If $2\ell_1-\kappa b^2\neq 0$, then we have $c_2=0$, and the solution is
\begin{gather*}
ds^2= -f dt^2 +\frac{dr^2}{f}+r^2d\Omega^2, \quad  f=1-\frac{R_s}{r},
\\
b_t = b, \quad b_r =\frac{b\sqrt{R_s r}}{r- R_s},
\end{gather*}
where $b$ and $R_s$ are free parameters.
Despite the entirely different derivation processes, the structure of the solution for $\lambda = \xi/2$ is identical to that for $\lambda = -\xi/2$ discussed in Sec.~\ref{sec:sol1}.

\subsubsection{\texorpdfstring{Case of $\lambda\neq0$ and $\lambda\neq \pm \xi/2$}{Case of lambda!=0 and lambda != ± xi/2}}

If $Q\neq 0$, $\ell_1\neq 0$, $\ell_1+2\ell_2\neq0$, and $\ell_1-2\ell_2\neq0$,
to ensure that $\mathcal{E}_{\theta\theta}$ is identically zero, we have
\begin{equation*}
(\ell_1+\ell_2)s=1, \quad s=c_1^2,
\end{equation*}
and the only solution is $s=c_1=1$ and $\ell_1+\ell_2=1$.
In this case, only solutions with \emph{timelike} VEVs exist.
The relation $\ell_1+\ell_2=1$ means that $b$ is a fixed value as
\begin{equation*}
b=\frac{1}{\sqrt{\lambda+\xi}}.
\end{equation*}
Besides, we have one more requirement $c_2{}^2 (2 \ell_1-b^2 \kappa )=0$.
If $2 \ell_1-b^2$ or $\xi=\kappa/2$, the solution is
\begin{gather*}
ds^2=-f dt^2+\frac{dr^2}{f}+r^2 d\Omega^2, \quad f= 1-\frac{R_s}{r}+\frac{Q}{r^p}, \quad p=\frac{4\lambda}{\xi+2\lambda},
\\
b_t = \frac{1}{\sqrt{\lambda+\xi}}+\frac{S}{r},\quad b_r =\frac{\sqrt{b_t^2-b^2 f}}{f},
\end{gather*}
where $R_s$, $Q$ and $S$ are free parameters.
If $c_2=0$, the solution is
\begin{gather*}
ds^2=-f dt^2+\frac{dr^2}{f}+r^2 d\Omega^2, \quad f= 1-\frac{R_s}{r}+\frac{Q}{r^p}, \quad p=\frac{4\lambda}{\xi+2\lambda},
\\
b_t = \frac{1}{\sqrt{\lambda+\xi}},\quad b_r =\frac{\sqrt{b_t^2-b^2 f}}{f},
\end{gather*}
where $R_s$, $Q$ and $S$ are free parameters.
As the value of $p$ is governed by the ratio $\lambda/\xi$, varying this ratio yields distinct classes of solutions. 
A notable example occurs at $\lambda = -\xi/4$ (where $p = -2$), in which the metric effectively recovers the Schwarzschild-(anti-)de Sitter geometry.

If $Q=0$, the constraints on the parameters become identical to those in the cases of $\lambda = \pm \xi/2$, and the resulting solutions also coincide with these two scenarios. We shall not pursue the details further here.

\subsection{Generalized Solutions with an Arbitrary Function}

In Ref.~\cite{Zhu:2025fiy}, we find that for the model with $\lambda=0$, if $\xi=\kappa/2$, $b^2=2/\kappa$, and the VEV of $B_\mu$ is timelike, then the solution can be parameterized by an arbitrary function.
Here we shall demonstrate that a similar situation also arises in the case where $\lambda \neq 0$.

By a suitable choice of coordinates, here we set $R(r)=r$.
To satisfy the constraint $B_\mu B^\mu=-b^2$, we use the field configuration as 
\begin{equation}
b_t = b \sqrt{G(r)} U(r), \quad b_r = b\frac{\sqrt{U(r)^2-1}}{\sqrt{H(r)}},
\end{equation}
where $U(r)$ is a function to be determined.
By analogy with the derivation in Ref.~\cite{Zhu:2025fiy}, and based on the results obtained in the previous section, the parametric conditions for the emergence of this situation are given by
\begin{equation*}
    \ell_1+\ell_2\equiv(\xi+\lambda)b^2=1, \quad \xi=\frac{\kappa}{2}. 
\end{equation*}
We shall demonstrate that, under the above parametric conditions, the solution can indeed be expressed in terms of an arbitrary function.

After replacing $b$ with $\sqrt{2\ell_1/\kappa}$, and replacing $\ell_2$ with $1-\ell_1$, we can obtain the equations of motions as follows
\begin{align*}
\mathcal{E}_{r}=&
\left(2-5 \ell_1\right) r^2 H U G'^2
+r G \left(2 \left(3 \ell_1-2\right) r H U G''
+G' \left(\left(3 \ell_1-2\right) U \left(r H'+4 H\right)
+4 \ell_1 r H U'\right)\right)
\\&
+4 G^2 \left(\ell_1 r \left(\left(r H'+4 H\right) U'+2 r H U''\right)+2 \left(\ell_1-1\right) U \left(r H'+H-1\right)\right),
\\
\mathcal{E}_r=&
-\left(\ell_1-2\right) r^2 H G'^2
+r G \left(2 \left(\ell_1-2\right) r H G''
+G' \left(\left(\ell_1-2\right) r H'
+8 \left(\ell_1-1\right) H\right)\right)
\\&
+4 G^2 \left(\left(\ell_1-2\right) r H'+2 \left(\ell_1-1\right) (H-1)\right),
\\
\mathcal{E}_{tt}=&
G' \left(r^2 G H' \left(4 \left(\ell_1-1\right) U^2-\ell_1+2\right)+4 \ell_1 r^2 G H U U'+4 r G H \left(\left(5 \ell_1-4\right) U^2-2 \ell_1+2\right)\right)
\\&
+2 r^2 G H G'' \left(4 \left(\ell_1-1\right) U^2-\ell_1+2\right)
+r^2 H G'^2 \left(\left(4-6 \ell_1\right) U^2+\ell_1-2\right)
+8 \ell_1 r^2 G^2 H U U''
+16 \ell_1 r G^2 H U U'
\\&
+H' \left(4 \ell_1 r^2 G^2 U U'+4 r G^2 \left(\left(3 \ell_1-4\right) U^2-\ell_1+2\right)\right)
+8 \left(\ell_1-1\right) G^2 (H-1) \left(2 U^2-1\right),
\\
\mathcal{E}_{rr}=&
-2 r^2 G H G'' \left(\ell_1+4 U^2-2\right)
+G' \left(r^2 (-G) H' \left(\ell_1+4 U^2-2\right)-4 \ell_1 r^2 G H U U'+4 r G H \left(\ell_1 \left(3 U^2-2\right)-4 U^2+2\right)\right)
\\&
+r^2 H G'^2 \left(2 \left(\ell_1+2\right) U^2+\ell_1-2\right)
+H' \left(4 r G^2 \left(\left(\ell_1-4\right) U^2-\ell_1+2\right)-4 \ell_1 r^2 G^2 U U'\right)
\\&
-8 \ell_1 r^2 G^2 H U U''
-16 \ell_1 r G^2 H U U'
+8 \left(\ell_1-1\right) G^2 (H-1) \left(2 U^2-1\right)
\\
\mathcal{E}_{\theta\theta}=&
2 \left(\ell_1-2\right) r^2 G H G''
+G' \left(\left(\ell_1-2\right) r^2 G H'-8 \ell_1 r^2 G H U U'+8 r G H \left(-\ell_1 \left(U^2-1\right)-1\right)\right)
\\&
-r^2 H G'^2 \left(\ell_1 \left(2 U^2+1\right)-2\right)
+4 \left(\ell_1-2\right) r G^2 H'
-8 \ell_1 r^2 G^2 H U'^2
-16 \ell_1 r G^2 H U U'
\\&
-8 G^2 \left(\ell_1 H \left(U^2-1\right)+H-1\right)
\\
\mathcal{E}_{\phi\phi}=&\sin^2(\theta)\mathcal{E}_{\theta\theta},
\\
\mathcal{E}_{tr}\propto&\mathcal{E}_{r}.
\end{align*}
Inspired by Ref.~\cite{Zhu:2025fiy}, we assume that the relation of $U(r)$ is given by
\begin{equation}
U(r)=\frac{ 1}{r\sqrt{G(r)}}\left(c_1+\int \sqrt{\frac{G(r)}{H(r)}}dr\right),\label{eq:U}
\end{equation}
where $c_1$ is a constant.
Substituting the relation of $U(r)$ into the equations of motion, we find that all of the equations reduce to a single one as
\begin{align}
&2 \left(\ell_1-2\right) r^2 G H G''
+G' \left(\left(\ell_1-2\right) r^2 G H'
+8 \left(\ell_1-1\right) r G H\right)
\nonumber\\&
-\left(\ell_1-2\right) r^2 H G'^2
+4 \left(\ell_1-2\right) r G^2 H'
+8 \left(\ell_1-1\right) G^2 (H-1)=0. \label{eq:GH}
\end{align}
Therefore, given an arbitrary function $G(r)$, one can successively determine $H(r)$ from \eq{eq:GH} and $U(r)$ from \eq{eq:U}, thereby constructing a complete solution for the system.

One may impose additional arbitrary constraints to single out a specific solution.
For example, if we require $G(r)=H(r)$, then the solution of \eq{eq:GH} is
\begin{equation*}
G(r)=H(r) = 1-\frac{R_s}{r}+Q r^{-\frac{4(\ell_1-1)}{\ell_1-2}}.
\end{equation*}
Since with $\ell_1+\ell_2=1$, we have
\begin{equation*}
p=\frac{4\lambda}{\xi+2\lambda}=\frac{4\ell_2}{\ell_1+2\ell_2}=\frac{4(\ell_1-1)}{\ell_1-2},
\end{equation*}
and 
\begin{equation*}
b_t = b \sqrt{G(r)} U(r) = b+\frac{c_2}{r}.
\end{equation*}
In other words, if we impose the condition $G(r)=H(r)$, the solution reduces to the case discussed in the preceding section.

\section{Summary of the Solutions}\label{sec:solutions}

In this section, we summarize the solutions obtained in the last section.
The solutions are classified into two classes: Class I solutions with the condition $b_r\equiv0$, and Class II solutions with the condition $\xi R^{r}_r+\lambda R \equiv 0$.

The solution of $b_\mu$ in Class I solutions are all \emph{timelike}, and the solutions are shown as follows:

\begin{description}

\item[Soluition I] With a fixed VEV $b=\frac{1}{\sqrt{\lambda}}$, the solution is
\begin{gather*}
ds^2=-dt^2+\left(1-\frac{R_s}{r}\right )^{-1} dr^2+r^2d\Omega^2, \\
b_t = \frac{1}{\sqrt{\lambda}}, \quad b_r = 0,
\end{gather*}
where $R_s$ is a free parameter.

\item[Solution II] With a fixed VEV $b=\sqrt{\frac{2}{\kappa+2\lambda}}$, the solution is
\begin{gather*}
ds^2=-\left(\frac{r}{R_s}\right)^{\frac{2(2-\ell)}{\ell}} dt^2+\frac{4}{\ell^2}dr^2+r^2 d\Omega^2, \quad \ell=\frac{2\xi}{\kappa+2\lambda}
\\
\quad b_t = \sqrt{\frac{2}{\kappa+2\lambda}} \left(\frac{r}{R_s}\right)^{2/\ell-1},
\quad b_r = 0,
\end{gather*}
where $R_s$ is a free parameter.
This solution can be interpreted as the result of substituting the constraint $B_\mu B^\mu = -b^2$ into the action and subsequently redefining the parameter $\kappa$ by
\begin{equation*}
\frac{1}{2\kappa'} R = \frac{1}{2\kappa}(R+\lambda B_\mu B^\mu R)=\frac{1-\ell_2}{2\kappa} R,
\end{equation*}
and replace $\kappa\to\kappa'=\kappa/(1-\ell_2)$ in the solution obtained in Ref.~\cite{Li:2025rjv}.

\item[Solution III] With a fixed VEV $b=\sqrt{\frac{2}{\kappa+2\lambda}}$, the solution is
\begin{gather*}
ds^2= -\left(1-\frac{R_s}{r}\right)^{2-\ell} dt^2 
+\left(1-\frac{R_s}{r}\right)^{-2+\ell} dr^2
+\left(1-\frac{R_s}{r}\right)^{\ell}r^2d\Omega^2, \quad \ell=\frac{2\xi}{\kappa+2\lambda}
\\
b_t =\sqrt{\frac{2}{\kappa+2\lambda}} \left(1-\frac{R_s}{r}\right)^{1-\ell/2},
\quad b_r=0,
\end{gather*}
where $R_s$ is a free parameter.
This solution can also be interpreted as the modification of the solution in Ref.~\cite{Li:2025rjv} by replacing $\kappa$ with the effective $\kappa'$.

\end{description}

Class II solutions are shown as follows:

\begin{description}

\item[Solution IV] If the parameters $\lambda$, $\xi$ are general, we have the solution
\begin{gather*}
ds^2= -f dt^2 +\frac{dr^2}{f}+r^2d\Omega^2, \quad  f=1-\frac{R_s}{r},
\\
b_t = b, \quad b_r =\frac{b\sqrt{R_s r}}{r- R_s},
\end{gather*}
where $R_s$ is a free parameter.
In the vector-tensor theory with $V\equiv 0$, $b$ is also a free parameter.

\item[Solution V] If the parameters satisfy the constraint  $\xi=\kappa/2$, the \textbf{Solution IV} can be extended as
\begin{gather*}
ds^2=-f dt^2+\frac{dr^2}{f}+r^2 d\Omega^2, \quad f= 1-\frac{R_s}{r},
\\
b_t = b+\frac{Q}{r},\quad b_r =\frac{\sqrt{b_t^2-b^2 f}}{f},
\end{gather*}
where $Q$ and $R_s$ are free parameters.
In the vector-tensor theory with $V\equiv 0$, $b$ is also a free parameter.

\item[Solution VI] If the parameters satisfy the constraint $\lambda\neq\pm \xi/2$, with a fixed VEV $b=\frac{1}{\sqrt{\lambda+\xi}}$, the solution is 
\begin{gather*}
ds^2=-f dt^2+\frac{dr^2}{f}+r^2 d\Omega^2, \quad f= 1-\frac{R_s}{r}+\frac{Q}{r^p}, \quad p=\frac{4\lambda}{\xi+2\lambda},
\\
b_t = \frac{1}{\sqrt{\lambda+\xi}},\quad b_r =\frac{\sqrt{b_t^2-b^2 f}}{f},
\end{gather*}
where $R_s$ and $Q$ are free parameters.

\item[Solution VII] If the parameters satisfy the constraint $\lambda\neq\pm \xi/2$ and $\xi = \kappa /2$, with a fixed VEV $b=\frac{1}{\sqrt{\lambda+\xi}}$, the \textbf{Solution VI} can be extended as
\begin{gather*}
ds^2=-f dt^2+\frac{dr^2}{f}+r^2 d\Omega^2, \quad f= 1-\frac{R_s}{r}+\frac{Q}{r^p}, \quad p=\frac{4\lambda}{\xi+2\lambda},
\\
b_t = \frac{1}{\sqrt{\lambda+\xi}}+\frac{S}{r},\quad b_r =\frac{\sqrt{b_t^2-b^2 f}}{f},
\end{gather*}
where $R_s$, $Q$ and $S$ are free parameters.

\item[Solution VIII] If the parameters satisfy the constraint $\xi=0$, then the coupling $B_\mu B_\nu R^{\mu\nu}$ vanishes, and the solution is a Reissner–Nordstr{\"o}m solution 
\begin{gather*}
ds^2=-f dt^2+\frac{dr^2}{f}+r^2 d\Omega^2, \quad f= 1-\frac{R_s}{r}+\frac{Q^2}{r^2},
\\
b_t= P + \sqrt{\frac{2(1-\lambda s b^2)}{\kappa}}\frac{Q}{r},
\quad b_r  =\frac{\sqrt{b_t^2-sb^2 f}}{f},
\end{gather*}
where $R_s$ and $Q$ are free parameters.
This solution can be viewed as the result of performing a gauge transformation on the vector field of the Reissner–Nordstr{\"o}m solution, followed by a redefinition of the parameter $\kappa$ as $\kappa'=\frac{\kappa}{1-\lambda s b^2}=\frac{\kappa}{1-s\ell_2}$.

\item[Solution IX] If the parameters satisfy the constraint $\xi=\kappa/2$, for a fixed VEV $b=\frac{1}{\sqrt{\lambda+\xi}}$, there exists a specific solution, which can be formulated with an arbitrary functional degree of freedom, as
\begin{gather*}
ds^2=-G(r)dr^2+\frac{ds^2}{H(r)}+r^2d\Omega^2,
\\
b_t(r) = \frac{b}{r}\left(c_1+\int \sqrt{\frac{G(r)}{H(r)}}dr\right),
\quad
b_r(r) =\sqrt{\frac{b_t(r)^2-b^2 G(r)}{G(r)H(r)}},
\end{gather*}
provided that the following conditions are satisfied:
\begin{align*}
&2 \left(\ell_1-2\right) r^2 G H G''
+G' \left(\left(\ell_1-2\right) r^2 G H'
+8 \left(\ell_1-1\right) r G H\right)
\nonumber\\&
-\left(\ell_1-2\right) r^2 H G'^2
+4 \left(\ell_1-2\right) r G^2 H'
+8 \left(\ell_1-1\right) G^2 (H-1)=0,
\end{align*}
where $\ell_1=\xi b^2 = \frac{\xi}{\lambda+\xi}=\frac{\kappa}{2\lambda+\kappa}$.
If we impose an additional condition $G(r)=H(r)$, we can obtain the same metric solutions as \textbf{Solution IV} and \textbf{VI}.
The relation between $G(r)$ and $H(r)$ is precisely the condition $\xi R^{r}_r+\lambda R = 0$ that must be satisfied by the Class II solutions.
An analysis in Appendix~\ref{sec:horizon} shows that if $G(r)$ possesses a first-order root at $r_0$, this generally implies that $H(r)$ also exhibits a simple zero at the same location. This suggests that, starting from this solution, one can construct an infinite family of black hole solutions characterized by an event horizon.
A particularly interesting case arises when $\lambda = -\xi/2$, where the vector field $B_\mu$ couples directly to the Einstein tensor $G_{\mu\nu}$.
While this configuration corresponds to the $\beta = 1/4$ case in Ref.~\cite{Chagoya:2016aar}, the general solution presented here was not identified in their work.
In this special case, $b=2/\sqrt{\kappa}$, and the relation between $G(r)$ and $H(r)$ is simple:
\begin{equation}
H(r)=\frac{G(r)}{G(r)+r G'(r)}.
\end{equation}

\end{description}

\section{Thermodynamics of the new black hole solutions}\label{sec:thermodynamics}

As a cornerstone of black hole physics, black hole thermodynamics offers a unique probe of quantum phenomena in curved spacetime~\cite{Hawking:1975vcx, Gibbons:1976ue, Bardeen:1973gs, Hawking:1976de}.
As point out in Refs.~\cite{An:2024fzf}, the Wald entropy formula~\cite{Wald:1993nt} is invalid in the Bumblebee case, which reads
\begin{equation}
S_{W}=-2\pi\int d^{2}x\sqrt{-\gamma}\frac{\partial L}{\partial R_{\mu\nu\rho\sigma}}\epsilon_{\mu\nu}\epsilon_{\rho\sigma}.
\end{equation}
The reason is the divergent behavior of the bumblebee field at the horizon.
For all of the Class II solutions, the Wald entropy formula gives
\begin{equation}
S_W=\frac{\mathcal{A}}{4G}\left(1-s\left(\ell_2+\frac{1}{2}\ell_1\right)\right),
\end{equation}
where $\mathcal{A}$ is the area of the horizon, $\ell_1=\xi b^2$ and $\ell_2=\lambda b^2$.
However, for all of the Class II solutions, the original Iyer–Wald formalism suggests that
\begin{equation}
S=\frac{\mathcal{A}}{4G}\left(1-s\left(\ell_2+\ell_1\right)\right).
\end{equation}
In the following, we shall employ the original Iyer–Wald formalism~\cite{Wald:1993nt, Iyer:1994ys, Iyer:1995kg} to analyze the thermodynamics of the new solutions and show the above result.
For notational simplicity, we set $G = 1$ in our thermodynamic calculations, and as a consequence $\kappa=8\pi$.

\subsection{A review of Iyer–Wald formalism}

For a gravity theory in 4d spacetime, the Lagrangian density is
$\mathbf{L}=L\boldsymbol{\epsilon}$, 
where $\boldsymbol{\epsilon}$ is the volume 4-form and $L$ is the Lagrangian in the action.
Varying the fields $\Phi\equiv \{g_{ab}, B_a\}$ gives
\begin{equation}
\delta\mathbf{L}=\mathbf{E}[\Phi]\delta\Phi+\mathrm{d}\boldsymbol{\Theta}[\Phi,\delta\Phi], \label{eq:dL}
\end{equation}
where $\mathbf{E}[\Phi]=0$ is the equation of motion of the fields, and $\boldsymbol{\Theta}[\Phi,\delta\Phi]$ is the presymplectic potential 3-form.
Consider the variation to be diffeomorphism $\delta_{\xi}\Phi=\mathcal{L}_{\xi}\Phi$, we find
\begin{equation}
\delta_\xi\mathbf{L}=\mathcal{L}_\xi\mathbf{L}=\xi\cdot\mathrm{d}\mathbf{L}+\mathrm{d}(\xi\cdot\mathbf{L})=\mathrm{d}(\xi\cdot\mathbf{L}).
\end{equation}
From this, \eq{eq:dL} becomes
\begin{equation}
\mathrm{d}(\xi\cdot\mathbf{L})=\mathbf{E}[\Phi]\mathcal{L}_{\xi}\Phi+\mathrm{d}\boldsymbol{\Theta}[\Phi,\mathcal{L}_{\xi}\Phi].
\end{equation}
It is easily seen that a Noether current 3-form can be defined by
\begin{equation}
\mathbf{J}_\xi=\mathbf{\Theta}[\Phi,\mathcal{L}_\xi\Phi]-\xi\cdot\mathbf{L},\label{eq:Jdef}
\end{equation}
since $\mathrm{d} \mathbf{J}_\xi=-\mathbf{E}[\Phi]\mathcal{L}_\xi\Phi$ which means that $\mathbf{J}_\xi$ is a closed form when equation of motions are satisfied.
As the closed form must be locally exact, there exists a two form $\mathbf{Q}_\xi$, such that $\mathbf{J}_\xi=\mathrm{d} \mathbf{Q}_\xi$.
A symplectic current can be constructed from $\mathbf{\Theta}$ as
\begin{equation}
\boldsymbol{\omega}(\mathbf{\Phi},\delta\mathbf{\Phi},\mathcal{L}_{\xi}\mathbf{\Phi})=\delta(\mathbf{\Theta}(\mathbf{\Phi},\mathcal{L}_{\xi}\mathbf{\Phi}))-\mathcal{L}_{\xi}(\mathbf{\Theta}(\mathbf{\Phi},\delta\mathbf{\Phi})).
\end{equation}
By doing the variation to \eq{eq:Jdef}, we have
\begin{equation}
\begin{aligned}
\delta\mathrm{d}\mathbf{Q}_{\xi}
&=\delta \mathbf{J}_\xi=\delta[\boldsymbol{\Theta}(\Phi,\mathcal{L}_{\xi}\Phi)]-\xi\cdot\delta\mathbf{L}
\\&
=\delta[\boldsymbol{\Theta}(\Phi,\mathcal{L}_\xi\Phi)]-\mathcal{L}_\xi[\boldsymbol{\Theta}[\Phi,\delta\Phi]]+d(\xi\cdot\boldsymbol{\Theta}[\Phi,\delta\Phi]).
\end{aligned}
\end{equation}
So we have the relation
\begin{equation}
\boldsymbol{\omega}(\mathbf{\Phi},\delta\mathbf{\Phi},\mathcal{L}_{\xi}\mathbf{\Phi})=
\mathrm{d}\left(\delta\mathbf{Q}_\xi-\xi\cdot\mathbf{\Theta}[\Phi,\delta\Phi]\right).
\end{equation}
By integrating above formula, we find
\begin{equation}
\int_{c}\boldsymbol{\omega}(\mathbf{\Phi},\mathbf{\Phi},\mathcal{L}_{\xi}\mathbf{\Phi})=\int_{\Sigma}(\delta \mathbf{Q}-i_{\xi}\mathbf{\Theta})=\delta H_{\infty}-\delta H_{+},
\end{equation}
where $c$ denotes Cauchy surface and $\Sigma$ is its boundary, which has two pieces, one is at asymptotic infinity and the other is at the horizon, and also $H$ is defined to be
\begin{equation}
\delta H_{\infty}=\int_{\infty}(\delta \mathbf{Q}-i_{\xi}\mathbf{\Theta}), \quad
\delta H_{+}=\int_{r_h}(\delta \mathbf{Q}-i_{\xi}\mathbf{\Theta}).
\end{equation}
When $\xi$ is a killing vector, $\mathcal{L}_{\xi}=0$, $\boldsymbol{\omega}(\mathbf{\Phi},\delta\mathbf{\Phi},\mathcal{L}_{\xi}\mathbf{\Phi})$ vanishes, which gives the relation 
\begin{equation}
\delta H_{\infty}=\delta H_{+}.
\end{equation}
The thermodynamic first law is the consequence of this formula.
For spherical symmetric case without angular momentum, $\delta H_{+}$ is identified as the variation of total spacetime energy which is defined at asymptotic infinity. 
It is called canonical Hamiltonian which will reduce to ADM Hamiltonian in Einstein gravity case. 
Meanwhile, $\delta H_{+}$ is identified as $T\delta S$ which is defined at black hole killing horizon, the entropy will have corrections compared to Bekenstein area law.

For the action~(\ref{eq:action}), the variation of the action gives the presymplectic potential as
\begin{equation}
\mathbf{\Theta}[\Phi,\delta\Phi]_{bcd}
=\varepsilon_{abcd}\Big(2E_{R}{}^{aefh}\nabla_{h}\delta g_{ef}-2\left(\nabla_{h}E_{R}{}^{aefh}\right)\delta g_{ef}-B^{ae}\delta B_{e}\Big),
\end{equation}
where $E_R^{abcd}=\frac{\partial L}{\partial R_{abcd}}$.
The Noether charge is
\begin{equation}
(\mathbf{Q}_\xi)_{cd}=\varepsilon_{abcd}\left(-E_R{}^{abef}\nabla_e\xi_f-2\xi_e\nabla_fE_R{}^{abef}-\frac{1}{2}B^{ab}B_f\xi^f\right).
\end{equation}
$E_R^{abcd}$ can be obtained by the following identities
\begin{equation}
\frac{\partial R}{\partial R_{abcd}}=\frac{1}{2}(g^{ac}g^{bd}-g^{ad}g^{bc}),
\end{equation}
\begin{equation}
\frac{\partial R_{ef}}{\partial R_{abcd}}=X_{ef}^{abcd}=\frac{1}{8}\left(
g^{ac}(\delta_e^d \delta_f^b + \delta_e^b\delta_f^d)
+g^{bd}(\delta_e^c\delta_f^a+\delta_e^a\delta_f^c)
-g^{ad}(\delta_e^c\delta_f^b+\delta_e^b\delta_f^c)
-g^{bc}(\delta_e^d\delta_f^a+\delta_e^a\delta_f^d)
\right).
\end{equation}

\subsection{Thermodynamics on specific solutions}

All of the black hole solutions can be parameterized as
\begin{gather*}
ds^2=-G(r)dr^2+\frac{ds^2}{H(r)}+r^2d\Omega^2,
\\
b_\mu = (b_t(r), b_r(r),0,0).
\end{gather*}
As a consequence, the variation of the fields is
\begin{gather*}
\delta g_{\mu\nu}=\mathrm{diag}\left(-\delta G(r), -\frac{\delta H(r)}{H(r)^2},0,0\right),
\\
\delta b_\mu=\left(\delta b_t(r), \delta b_r(r),0,0\right).
\end{gather*}
For the killing vector $\xi^\mu=\partial_t$, only the $\theta\phi$ (and $\phi\theta$) components of $\delta \mathbf{Q}$ and $i_{\xi}\mathbf{\Theta}$ are non-vanishing.
To simplify the notation, we denote
\begin{equation}
(\delta \mathbf{Q})_{\theta\phi}\equiv \sqrt{-g} \delta Q, \quad
(i_{\xi}\mathbf{\Theta})_{\theta\phi} \equiv \sqrt{-g} i_{\xi}\Theta.
\end{equation}
In this notation and the field configuration, we have

\begin{equation}
\begin{aligned}
\delta Q-i_{\xi}\Theta
=&
-\frac{(2 \lambda +\xi )H b_t^2  }{2 \kappa  G^2} \delta G'
-\frac{(2 \lambda +\xi )H b_r^2  }{2 \kappa }\delta H'
-\frac{(\kappa -2 \lambda -\xi )H b_t }{\kappa  G}\delta b_t' 
-\frac{(2 \lambda +\xi )H^2 b_r   }{\kappa }\delta b_r'
\\&
+\frac{ H b_t  \left(2(\kappa -2 (2 \lambda +\xi )) G b_t' +5(2 \lambda +\xi )  G' b_t \right)}{4 \kappa  G^3}\delta G
+\frac{ (2 \lambda +\xi ) H \left(2 G b_t'-3 b_t G'\right)}{2 \kappa  G^2}\delta b_t 
\\&
+\frac{ H \left(b_r \left((2 \lambda +\xi ) r H  G'-2 G \left( (2 \lambda +\xi ) r H'+2 \xi  H\right)\right)-2 r G H (2 \lambda +\xi ) b_r'\right)}{2 \kappa  r G}\delta b_r
\\&
+\frac{1}{4 \kappa  r G^2}\Big(
G \left(r b_r^2 H (2 \lambda +\xi ) G'-G \left(6 r b_r H (2 \lambda +\xi ) b_r'+b_r^2 \left(r (2 \lambda +\xi ) H'+4 H (\lambda +2 \xi )\right)+4\right)\right)
\\&\quad\quad\quad\quad\quad
+2 r b_t G b_t' (-\kappa +2 \lambda +\xi )
+b_t^2 \left(4 \lambda  G-r (2 \lambda +\xi ) G'\right)
\Big)\delta H.\label{eq:QTheta}
\end{aligned}
\end{equation}
In what follows, we shall demonstrate the thermodynamics of these black hole solutions, specifically focusing on the non-trivial regime where $\lambda \neq 0$ and $\xi \neq 0$.

\subsubsection{Solution IV}

In this solution, denote $R_s=2M$, we have
\begin{gather*}
G = H = f = 1-\frac{2M}{r}, \quad \delta G=\delta H=\delta f=-\frac{2\delta M}{r},
\\
b_t=b,\quad b_r= \frac{b}{f}\sqrt{1-f},\quad
\delta b_t=0, \quad
\delta b_r = \frac{b (f-2)}{2 f^2\sqrt{1-f}}\delta f.
\end{gather*}
Substitute it into \eq{eq:QTheta}, we have
\begin{equation}
\delta Q-i_{\xi}\Theta = -\frac{1-b^2(\lambda+\xi)}{\kappa r}\delta f =(1-\ell_1-\ell_2)\frac{\delta M}{4\pi r^2}, \label{eq:rel4}
\end{equation}
where $\ell_1=\xi b^2$ and $\ell_2=\lambda b^2$.
The integration at asymptotic infinity gives 
\begin{equation}
\delta H_\infty =\delta E= (1-\ell_1-\ell_2)\delta M.
\end{equation}
Also, evaluating this on the horizon gives
\begin{equation}
\delta H_{+}=T\delta S = \frac{1}{2} (1-\ell_1-\ell_2) \delta r_h.
\end{equation}
The temperature of a black hole is easily calculated either by using the Euclidean method or the quantum tunneling method~\cite{Gomes:2018oyd} as
\begin{equation}
T=\frac{\sqrt{G'(r_h)H'(r_h)}}{4\pi}=\frac{1}{4\pi r_h}.
\end{equation}
The above results imply that we should define the thermodynamic variables as follows
\begin{equation}
E = (1-\ell_1-\ell_2)M, \quad S=\frac{\mathcal{A}}{4}(1-\ell_1-\ell_2),
\end{equation}
where $\mathcal{A}$ is the area of the horizon.
The energy, temperature, and entropy still satisfy the following Smarr relation
\begin{equation}
E=2TS.
\end{equation}
Although the metric is still the Schwarzschild metric, the thermodynamics is different.
Due to the presence of the background field $b_\mu$ and the non-minimal coupling, the entropy receives a correction by a factor of $1 - \ell_1 - \ell_2$.

\subsubsection{Solution V}
In this solution, the parameters satisfy the constraint $\xi=\kappa/2$.
With $Rs=2M$, the field configuration can be expressed as
\begin{gather*}
G = H = f = 1-\frac{2M}{r}, \quad \delta G=\delta H=\delta f=-\frac{2\delta M}{r},
\\
b_t=b+\frac{C}{r},\quad b_r =\frac{\sqrt{b_t^2-b^2 f}}{f},\quad
\delta b_t=\frac{\delta C}{r}, \quad
\delta b_r =\frac{b_t \delta b_t}{f\sqrt{b_t^2-b^2 f}}+\frac{(-2b_t^2+b^2 f)\delta f}{2f^2 \sqrt{b_t^2-b^2 f}}.
\end{gather*}
Substitute it into \eq{eq:QTheta}, we have
\begin{equation}
\delta Q-i_{\xi}\Theta = 
-\frac{1-b^2(\lambda+\xi)}{\kappa r}\delta f 
+\frac{(\kappa-2\xi)(C+b r)}{\kappa r^3}\delta C.
\end{equation}
Since $\xi=\kappa/2$, the variation of $b_t$ does not contribute, and it reduces to the same form as \eq{eq:rel4}.
As a consequence, the thermodynamics is the same as {Solution IV}.

\subsubsection{Solution VI and VII}


In these two solutions, we have
\begin{gather*}
G = H = f, \quad \delta G=\delta H=\delta f,
\\
b_t=b+\frac{C}{r},\quad b_r =\frac{\sqrt{b_t^2-b^2 f}}{f},\quad
\\
\delta b_t=\frac{\delta C}{r}, \quad
\delta b_r =\frac{b_t \delta b_t}{f\sqrt{b_t^2-b^2 f}}+\frac{(-2b_t^2+b^2 f)\delta f}{2f^2 \sqrt{b_t^2-b^2 f}}.
\end{gather*}
This differs from Solutions IV and V only in the form of $f$.
Substitute it into \eq{eq:QTheta}, we have
\begin{equation}
\delta Q-i_{\xi}\Theta = 
-\frac{1-b^2(\lambda+\xi)}{\kappa r}\delta f 
+\frac{(\kappa-2\xi)(C+b r)}{\kappa r^3}\delta C.
\end{equation}
In both solutions, since $b=1/\sqrt{\lambda+\xi}$, the term proportional to $\delta f$ vanishes.
For Solution VI, $C\equiv 0$, so the second term vanishes.
For Solution VI, since $\xi=\frac{\kappa}{2}$, the second term also vanishes.
Thus we have 
\begin{equation}
\delta Q-i_{\xi}\Theta \equiv0.
\end{equation}
This means $\delta H_{+}=\delta H_{\infty}=0$, and the entropy of this black hole is zero.

\subsubsection{Solution VIII}

Here we consider the special case $\xi=0$, in which the solution can be viewed as the result of performing a gauge transformation on the vector field of the RN solution, followed by a redefinition of the parameter $\kappa$ as $\kappa'=\frac{\kappa}{1-\lambda s b^2}=\frac{\kappa}{1-s\ell_2}$.
With $R_s=2M$, the field configuration is
\begin{gather*}
G = H = f = 1-\frac{2M}{r}+\frac{Q^2}{r^2}, \quad \delta G=\delta H=\delta f=-\frac{2\delta M}{r}+\frac{2Q\delta Q}{r^2},
\\
b_t= P + \sqrt{\frac{2(1-\lambda s b^2)}{\kappa}}\frac{Q}{r},
\quad b_r  =\frac{\sqrt{b_t^2-sb^2 f}}{f},
\\
\delta b_t= \delta P + \sqrt{\frac{2(1-\lambda s b^2)}{\kappa}}\frac{\delta Q}{r}, \quad
\delta b_r =\frac{b_t \delta b_t}{f\sqrt{b_t^2-sb^2 f}}+\frac{(-2b_t^2+sb^2 f)\delta f}{2f^2 \sqrt{b_t^2-sb^2 f}}.
\end{gather*}
Substitute the field configuration into \eq{eq:QTheta}, we have
\begin{equation}
\delta Q-i_{\xi}\Theta = -\frac{1-s\ell_2}{\kappa r}\delta f(r) -b_t(r)\delta b_t'(r).
\end{equation}
The integration at $r=\infty$ gives
\begin{equation}
\delta H_\infty = (1-s\ell_2)\delta M +{2}{\sqrt{\pi(1-s\ell_2)}}P\delta Q.
\end{equation}

Then we perform the integration at the outer horizon $r=r_{+}$.
From $f(r_+)=0$, we have $\delta f(r_+)=0$. 
The variation of $\delta f(r_+)$ has two contributions:
\begin{equation}
\delta f(r_+) = \frac{\partial f}{\partial r} \Big |_{r=r+} \delta r_+ + \delta f\Big |_{r=r+}, 
\end{equation}
so we have $\delta f(r_+)=-f'(r+)=-2\boldsymbol{\kappa}$, here $\boldsymbol{\kappa}=f'(r_+)/2$ is the surface gravity.
Hence, at the outer horizon, we have
\begin{equation}
(\delta Q-i_{\xi}\Theta)\Big|_{r=r+}=\frac{1-s\ell_2}{4\pi r_{+}^2}\frac{\boldsymbol{\kappa}}{2\pi}\delta(\pi r_+^2)
+\Big(
\frac{1-s\ell_2}{4\pi r_{+}^3}Q
+\frac{\sqrt{1-s\ell_2}}{2\sqrt{\pi}r_+^2}P
\Big)\delta Q.
\end{equation}
The integration at $r=r_+$ gives
\begin{equation}
\delta H_+ = (1-s\ell_2)\frac{\boldsymbol{\kappa}}{8\pi}\delta \mathcal{A}+(1-s\ell_2)\frac{Q}{r_+}\delta Q+{2}{\sqrt{\pi(1-s\ell_2)}}P\delta Q,
\end{equation}
where $\mathcal{A}=4 \pi r_+^2$ is area of the outer horizon.
Since $\delta H_\infty = \delta H_+$, the term proportional to $P\delta Q$ can be eliminated, and we obtain
\begin{equation}
(1-s\ell_2)\frac{\boldsymbol{\kappa}}{8\pi}\delta \mathcal{A}+(1-s\ell_2)\frac{Q}{r_+}\delta Q=(1-s\ell_2)\delta M.
\end{equation}
Hence we have
\begin{equation}
E=(1-s\ell_2)M,\quad S=\frac{\mathcal{A}}{4}(1-s\ell_2)=\frac{A}{4 G_{\rm eff}},
\end{equation}
where $G_{\rm eff}=\frac{\kappa'}{8\pi}=\frac{1}{1-s\ell_2}$.
The thermodynamically effective gravitational constant is consistent with the redefined effective gravitational constant of the RN solution.
It is worth noting that in this case, the entropy formula coincides with the Wald entropy formula, since $\ell_1=0$.

\subsubsection{Solution IX}

Lastly, we consider the case when Solution IX is a black hole solution.
From the relation
\begin{equation}
b_r(r) =\sqrt{\frac{b_t(r)^2-b^2 G(r)}{G(r)H(r)}},
\end{equation}
we have
\begin{equation}
\delta b_r=\frac{b^2  G^2 \delta H+2 b_t  G H \delta b_t-b_t^2 ( H\delta G+ G \delta H )}{2 G^{3/2} H^{3/2} \sqrt{b_t^2-b^2 G}}.
\end{equation}
Varying the relation
\begin{equation}
b_t(r) = \frac{b}{r}\left(C+\int \sqrt{\frac{G(r)}{H(r)}}dr\right),
\end{equation}
we obtain
\begin{equation}
\delta b_t = \frac{b}{r}\left(\delta C + \int\frac{H\delta G-G\delta H}{2H\sqrt{GH}} dr\right).
\end{equation}
Substitute these relations into \eq{eq:QTheta}, we have
\begin{equation}
\delta Q-i_{\xi}\Theta =
\frac{1}{\kappa r G^2}
\Big(
-(1-b^2(\lambda+\xi))G^2\delta H
+(\kappa-2\xi)b_t G H \delta b_t
+\frac{1}{2}(\kappa-2\xi)b_t^2 (G\delta H- H\delta G)
\Big).
\end{equation}
Since $b=1/\sqrt{\lambda+\xi}$ and $\xi=\kappa/2$ are satisfied in this solution, we have
\begin{equation}
\delta Q-i_{\xi}\Theta \equiv0.
\end{equation}
So if this solution is a black hole, then its entropy is zero.

\section{Discussion and Summary}\label{sec:discusion}

\subsection{Discussion on the solutions}

In what follows, we discuss these solutions.
\textbf{Solutions II} and \textbf{III} have been discussed in detail in Ref.~\cite{Li:2025rjv} and will not be elaborated upon here.

For Class I solutions, different from the case of $\lambda=0$, which is discussed in detail in Ref.~\cite{Li:2025rjv}, we find a novel solution as \textbf{Solution I}.
The Kretschmann scalar $K=R_{abcd}R^{abcd}=\frac{6R_s^2}{r^6}$ suggests that $r=R_s$ is not a singularity.
Since $g_{tt}=1$, there is no gravitational redshift from the perspective of an observer at infinity.
Compared with the standard Morris–Thorne metric~\cite{Morris:1988cz, Morris:1988tu}, this solution is evident to be a \emph{wormhole}, with redshift function $\Phi(r)\equiv 0$ and shape function $b(r)\equiv R_s$.
To see that, we can rewrite the metric in a new coordinate as
\begin{equation}
ds^2=-dt^2+dl^2 + r(l)^2d\Omega^2,
\end{equation}
where $l=\sqrt{r(r-R_s)}+R_s \mathrm{arctanh}\left(\sqrt{1-R_s/r}\right)$.
Near $r=R_s$, the relation between $l$ and $r$ can be approximated by $r\simeq \frac{l^2+4R_s^2}{4 R_s}$, and the metric is
\begin{equation}
ds^2\simeq -dt^2+dl^2+\Big(R_s+\frac{l^2}{4R_s}\Big)^2d\Omega^2.
\end{equation}
The metric admits an analytic extension into the $l < 0$ region, describing a traversable wormhole with a throat radius of $R_s$.
Unlike the case in GR, where wormholes are synonymous with energy condition violations, non-minimal coupling between the vector field and curvature can source wormhole geometries even in the bumblebee vacuum.
For the vector-tensor theory with $V\equiv0$ and $\lambda>0$, this wormhole solution always exists.
However, if the theory is a Bumblebee theory, the wormhole solution exists if and only if the condition $b^2 = 1/\lambda$ is satisfied.
A distinctive feature of Class II solutions, in contrast to the $\lambda = 0$ case~\cite{Zhu:2025fiy}, is that the vector field VEV is restricted to being either \emph{timelike} or \emph{lightlike}, except when the $B_\mu B_\nu R^{\mu\nu}$ coupling vanishes at $\xi = 0$.
This result is in alignment with analysis from a Hamiltonian perspective~\cite{BumblebeeV}, which demonstrates that SSB of a vector field restricts its VEV to being either timelike or lightlike.

Another novel result is that, for general parameters $\lambda$, $\xi$, and $b$, there always exists a solution (\textbf{Solution IV}) where the metric is of the Schwarzschild form.
This solution remained unidentified in Ref.~\cite{Chagoya:2016aar}, due to the restrictive assumptions placed on $b_t$.
We also note that \textbf{Solution V} coincides exactly with the solution reported in Ref~\cite{Chagoya:2016aar}.

For Class II solutions, when $b^2=1/(\xi+\lambda)$, we find totally new solutions (\textbf{Solution VI} and \textbf{VII}) where the metric is of the form $-g_{tt}={g_{rr}}^{-1}=1-R_s/r+Q/r^p$.
The interesting part is that the exponent $p$ is determined entirely by the ratio of the parameters $\xi$ and $\lambda$ as $p={4\lambda}/({\xi+2\lambda})$.
When $\xi=\pm 2\lambda$, the solutions reduce to the Schwarzschild metric.
When $\xi=0$, the solutions reduce to the RN solution presented as \textbf{Solution VIII}.
If the parameters $\xi$ and $\lambda$ can be chosen arbitrarily, the exponent $p$ can take any fractional value, or even be an irrational or transcendental number.
An interesting example is that when $\xi=-4\lambda$, the solution reduces to (A)dS solution.
If we want the term $Q/r^p$ to be suppressed by $R_s/r$, we have $p\leq1$, which gives the constraint $-1/2 \leq\lambda/\xi \leq 1/2$.
If $b^2=1/(\xi+\lambda)$ and $\xi=\kappa/2$ are satisfied simultaneously, similar to the special case $\lambda=0$, we find that the solution can be parameterized by an arbitrary function (\textbf{Solution IX}). 
This solution is a generalization of the special case $\lambda=0$ found in Ref.~\cite{Zhu:2025fiy}.
Such a scenario implies that the dynamical equations are degenerate and lose their ability to constrain the fields.
It seems that this specific set of parameters should be excluded from the physically viable parameter space, representing a singular boundary where the underlying theory lacks a well-defined dynamical structure.

It is worth discussing the thermodynamic aspects of the new black hole solutions.
For all of the solutions obtained in this work, the Iyer–Wald formalism shows that the entropy of the black hole can be expressed as a unified expression as
\begin{equation}
S=\frac{\mathcal{A}}{4G}\left(1-sb^2\left(\lambda+\xi\right)\right),
\end{equation}
where $\mathcal{A}$ is the area of the horizon.
The result is different from the result calculated directly from the Wald entropy formula, which gives
\begin{equation}
S_W=\frac{\mathcal{A}}{4G}\left(1-sb^2\left(\lambda+\xi/2\right)\right).
\end{equation}
The inconsistency between the calculated entropy and the Wald entropy was previously noted in the $\lambda = 0$ limit~\cite{An:2024fzf, Liu:2025oho}.
It is worth noting that for the solutions with $b^2=1/(\lambda+\xi)$ (\textbf{Solutions VI}, \textbf{VII} and \textbf{IX}), the entropy is identically zero.
This vanishing arises from a delicate cancellation among the gravity sector, the non-minimal coupling, and the contribution from the background field $b_\mu$.
This behavior suggests that these solutions reside in a degenerate sector of the theory, where the effective gravitational coupling is driven to zero by the vector field background.

\subsection{Why bumblebee theory}

For a system possessing multiple solutions, one might argue that unphysical branches should simply be discarded. However, a natural requirement for a theory describing the physical world is that such unphysical solutions should not arise mathematically in the first place, much as is the case in GR.
Based on this requirement, we argue that for a vector-tensor modified gravity theory to be viable, it should incorporate a potential that triggers a VEV, as seen in the Bumblebee model.
For clarity, we shall refer to the theory with $V \equiv 0$ as the ``free theory", to distinguish it from the Bumblebee theory.

For all of the solutions obtained in this work, they are also the solutions to the free theory.
The difference is that the only parameters are $\xi$ and $\lambda$, and $b$ is now a free parameter in the free theory.
The static spherically symmetric vacuum solutions to the free theory can also be classified into Class I and Class II.
Given that $b$ is a free parameter, these two branches of solutions are mathematically present. 
Nevertheless, one can easily show that Classes I and II are disconnected, as their intersection necessitates a flat Minkowski metric. 
This implies that for a realistic physical scenario, one must make a mutually exclusive choice between these two classes of solutions, rendering the other branch unphysical. 
Based on the aforementioned principles, the free theory appears theoretically untenable.
Furthermore, the free theory yields non-trivial black hole solutions with vanishing entropy~\footnote{In free theories, similar to the cancellation of the contribution from $P$ in \textbf{Solution VIII}, the variation $\delta b$ in the Iyer–Wald calculation yields no contribution.} (\textbf{Solutions VI}, \textbf{VII} and \textbf{IX}). On the grounds that such thermodynamic anomalies are unphysical, we argue that the free theory is theoretically disadvantaged.

Another argument is from a phenomenological perspective.
Consider a scenario where we adopt a Minkowski background and assume the vector field $B_\mu$ to be timelike. In this context, one can always perform a Lorentz transformation such that, in the new coordinate system, only the temporal component of $B_\mu$ remains non-vanishing.
By placing a static massive source in this frame, the resulting gravitational backreaction deforms the spacetime. This leads the metric and the $B_\mu$ field to evolve into the vacuum configurations characteristic of the free theory.
Since the spatial component of $B_\mu$ in the Minkowski background is zero, we expect the spatial component of $B_\mu$ in the deformed spacetime to also be zero.
That is, the solution is expected to be of Class I.
However, Class I solutions have a naked singularity, and they cannot smoothly evolve into the black hole solutions of Class II.
For the bumblebee theory, the situation is markedly different.
Although we have presented the vacuum solution for the bumblebee theory at the VEV, there can exist vacuum solutions away from the VEV~\cite{Bailey:2025oun}, and the solutions depend on the form of the potential.
For the above scenario, the spacetime and the $B_\mu$ first deform into a solution away from the VEV with $b_r=0$, then the SSB triggers, and it transforms into a Class II solution.

A final argument concerns the structure of the solution space.
In the minimally coupled Einstein-Maxwell theory, the presence of gauge symmetry implies that many solutions, though distinct in their mathematical form, are physically equivalent.
In non-minimally coupled theories, however, the absence of gauge symmetry transforms these former redundancies into true physical degrees of freedom. Consequently, the solution space encompasses a greater number of new degrees of freedom.
In the following, we present an illustrative example to further clarify this point.
We consider the theory with $\lambda=-\xi/2$, and the non-minimal coupling becomes $\frac{\xi}{2\kappa}B_\mu B_\nu G^{\mu\nu}$.
For general parameters, the bumblebee theory has only one solution (\textbf{Solution IV}), and the only free parameter is $R_s$.
However, in the case of the free theory, we find that the Class II solutions are determined by three independent parameters, even under the requirement of asymptotic flatness. Although an exact solution remains elusive, the solution obtained via asymptotic expansion at infinity is given by
\begin{equation*}
ds^2=-G(r)dt^2+\frac{1}{H(r)}dr^2+r^2d\Omega^2,\quad B_\mu=(b_t(r),b_r(r),0,0),
\end{equation*}
where
\begin{equation*}
\begin{aligned}
G=&
1-\frac{R_s}{r}
+\frac{ \kappa b_0^2  \left(q^2-1\right)^2 (\kappa -2 \xi ) R_s^2}{4  \left(\kappa  \left(2-3 b_0^2 \xi \right)+4 b_0^2 \xi ^2\right)r^2}
+\frac{ \kappa  \xi  b_0^4\left(q^2-1\right)^3 (\kappa -2 \xi )^2 R_s^3}{3  \left(\kappa  \left(2-3 b_0^2 \xi \right)+4 b_0^2 \xi ^2\right){}^2 r^3}+\cdots,
\\
H= &
1-\frac{R_s}{r}
-\frac{ \kappa b_0^2 \left(q^2-1\right)^2 (\kappa -2 \xi ) R_s^2}{ \left(\kappa  \left(6 b_0^2 \xi -4\right)-8b_0^2 \xi ^2\right)r^2}
+\frac{ \kappa b_0^2  \left(q^2-1\right)^2 (\kappa -2 \xi ) R_s^3 \left(\kappa  \left(b_0^2 \xi  \left(4
   q^2-1\right)-2\right)+4 b_0^2 \xi ^2 \left(1-2 q^2\right)\right)}{4  \left(\kappa  \left(2-3 b_0^2 \xi\right)+4 b_0^2 \xi ^2\right){}^2 r^3}+\cdots,
\\
b_t=&
b_0+\frac{b_0 \left(q^2-1\right) R_s}{2 r}
+\frac{b_0^3 \xi  \left(q^2-1\right)^2 (\kappa -2 \xi ) R_s^2}{4  \left(\kappa  \left(2-3 b_0^2 \xi \right)+4b_0^2 \xi ^2\right)r^2}
+\frac{b_0^3 \left(q^2-1\right)^3 (\kappa -2 \xi )^2 R_s^3 \left(\kappa  \left(3 b_0^2 \xi -2\right)+12 b_0^2 \xi
   ^2\right)}{48 \left(\kappa  \left(2-3 b_0^2 \xi \right)+4 b_0^2 \xi ^2\right){}^2 r^3 }+\cdots,
\\
b_r=&
q b_0\sqrt{\frac{R_s}{r}}
+b_0 \left(\frac{R_s}{r}\right)^{\frac{3}{2}}\frac{64 b_0^2 \xi ^3 q^2+8 \kappa 
   \xi  \left(q^2 \left(2-6 b_0^2 \xi \right)+q^4+1\right)-\kappa ^2 \left(q^2-1\right)^2 \left(b_0^2 \xi +2\right)}{16 \xi  q  \left(\kappa  \left(2-3 b_0^2
   \xi \right)+4 b_0^2 \xi ^2\right)}+\cdots.
\end{aligned}
\end{equation*}
The solution is presented by three free parameters $R_s$, $b_0$, and $q$, and it is easy to check that when $q=1$, the solution reduces to \textbf{Solution IV} in this work.
If the non-minimal coupling does not take the form of a contraction between the Einstein tensor and the vector field, the Class II solutions possess up to five independent free parameters.
While exact solutions with five parameters remain challenging to obtain, several examples of four-parameter solutions can be presented.
A first example is of the case $\lambda=0$, where a four-parameter exact solution is found in Ref.~\cite{Xu:2022frb}.
A second example is of the case $\xi=0$, and \textbf{Solution VII} in this work is a four-parameter ($R_s$, $P$, $Q$, and $b$) exact solution for the free theory.
A third example is also of the case $\xi=0$.
In this case, in the metric sector, besides $R_s$ and $Q$, there is one more parameter in the solution.
If this parameter is chosen to be a special one, the solution reduces to \textbf{Solution VII}.
We find that if this parameter is chosen to be another special one, we can obtain an exact asymptotically flat solution (see Appendix~\ref{sec:4sol}).
This solution consists of two distinct branches, and interestingly, these two branches can be spliced together to construct a complete and traversable  Morris–Thorne wormhole geometry. 
These diverse examples collectively demonstrate that without a symmetry-breaking potential, the free vector-tensor gravity is characterized by excessive freedom. This not only leads to a proliferation of free parameters but also results in an unconstrained solution space where distinct causal structures, ranging from black holes to traversable wormholes and naked singularities, can coexist and be constructed.
To achieve a physically constrained and predictive theory, the Bumblebee gravity framework is therefore preferred, as the inclusion of a potential effectively narrows the vast solution space.

In conclusion, the free vector-tensor gravity is characterized by excessive freedom. 
Without the constraint of a symmetry-breaking potential, the theory yields a proliferation of unphysical vacuum solutions and suffers from a lack of predictivity.
The Bumblebee framework is therefore preferred, as the potential effectively narrows the vast solution space into a physically consistent regime.

\subsection{Summary}

In this work, we have presented a systematic investigation of static spherically symmetric vacuum solutions within an extended bumblebee gravity model characterized by non-minimal kinetic-type couplings $B^{2}R$ and $B^{\mu}B^{\nu}R_{\mu\nu}$. 
By identifying the non-commutative nature between the variation of the action and the imposition of the VEV constraint, we have uncovered a significantly richer solution space than previously explored in simpler vector-tensor models.

For Class I solutions ($b_r \equiv 0$), we found two naked singularity solutions and a novel traversable wormhole solution.
Unlike General Relativity, this wormhole geometry is supported by the non-minimal coupling of the vector field rather than the violation of energy conditions by exotic matter.
For Class II solutions ($\xi R_r^r + \lambda R \equiv 0$), we derived a diverse set of black hole solutions.
Notably, we found that for general parameters, a Schwarzschild-like metric always exists, which remained unidentified in Ref.~\cite{Chagoya:2016aar}, due to the restrictive assumptions placed on $b_t$.
We also identified cases where the metric coefficients follow a power-law behavior $r^{-p}$, where the exponent $p$ is determined solely by the ratio of the coupling constants.
Furthermore, when $\xi=\kappa/2$, we found a solution that can be characterized by an arbitrary function, which indicates that the non-minimally coupled free vector-tensor theory is ill-defined when $\xi=\kappa/2$.

We also performed a thermodynamic analysis of these black hole solutions and found that the results obtained from Wald’s entropy formula are inconsistent with those derived from the Iyer-Wald formalism.
The thermodynamic analysis using the Iyer-Wald formalism reveals that the black hole entropy receives a consistent correction factor $1 - sb^2(\lambda+\xi)$ due to the Lorentz-violating background. Intriguingly, a subset of these new solutions (\textbf{Solutions VI}, \textbf{VII}, and \textbf{IX}) exhibits zero entropy, suggesting they reside in a degenerate sector of the theory where effective gravitational coupling vanishes.

Finally, we argued from both mathematical and phenomenological perspectives that the Bumblebee-type theory, where the vector field is constrained by a potential, is physically more viable than a ``free" vector-tensor theory. The presence of the potential effectively prunes unphysical branches and resolves the excessive freedom in the solution space, ensuring a more predictive framework for modified gravity. 
As a demonstration, we further provide a highly non-trivial traversable wormhole solution for the $\xi=0$ free theory in Appendix~\ref{sec:4sol}.
This work further deepens the insights into the landscape of vector-tensor coupling theories.
These results provide new insights into the potential role of Lorentz symmetry breaking in the strong-field regime and offer a broader range of templates for future gravitational wave and black hole shadow observations.


\section*{Acknowledgements}

This work was supported in part by the National Natural Science Foundation of China under Grant No.~12547101. HL was also supported by the start-up fund of Chongqing University under No.~0233005203009, and JZ was supported by the start-up fund of Chongqing University under No.~0233005203006.

\appendix

\section{Morris–Thorne wormhole metric}

The standard Morris–Thorne wormhole metric is~\cite{Morris:1988cz, Morris:1988tu}
\begin{equation}
ds^2=- e^{2\Phi(r)}dt^2+\frac{1}{1-b(r)/r}dr^2+r^2d\Omega^2,
\end{equation}
where $\Phi(r)$ is the redshift function and $b(r)$ is the shape function.

\section{Existence of horizon for Class II solutions}\label{sec:horizon}

In the following, we demonstrate that for the Class II vacuum solutions, if $r=r_0>0$ is a first-order root of $g_{tt}$, it necessarily corresponds to a first-order singularity of $g_{rr}$.
This implies that if $g_{tt} = 0$ admits a solution at $r=r_0$, then $r=r_0$ corresponds to an event horizon.

For the field configuration
\begin{equation*}
ds^2=-G(r)dt^2+\frac{1}{H(r)}dr^2+r^2d\Omega^2,
\end{equation*}
the equation for the condition $\xi R^{r}_r+\lambda R \equiv 0$ is
\begin{equation}
\alpha  r^2 H G'^2-r G \left(\alpha  r G' H'+2 H \left(\alpha  r G''+4 \lambda  G'\right)\right)-4 G^2 \left(\alpha  r H'+2 \lambda  (H-1)\right)=0,\label{eq:BGH}
\end{equation}
where $\alpha=2\lambda+\xi$.
If $\alpha=0$, implying that the vector field couples directly to the Einstein tensor, we have
\begin{equation*}
H(r)=\frac{G(r)}{G(r)+rG'(r)}.
\end{equation*}
If $r=r_0>0$ is a first-order root of $G(r)$, the $G(r_0)=0$ and $G'(r_0)\neq 0$, hence $r=r_0$ is also a first-order root of $H(r)$.
If $\alpha\neq 0$, then the solution for $H(r)$ from \eq{eq:BGH} is
\begin{equation*}
H(r)=P(r)\Big(c_1 + \frac{8\lambda}{\alpha} \int\frac{dr}{P(r)(4+X(r))r}\Big),
\end{equation*}
where
\begin{equation}
\begin{aligned}
P(r) =& \frac{G(r)^2}{(4G(r)+r G'(r))^2}\exp\Big(\frac{1}{\alpha}\int\frac{\alpha X(r)(2-X(r))-8\lambda(1+X(r))}{(4+X(r))r}dr\Big),\\
X(r) =& \frac{r G'(r)}{G(r)}.
\end{aligned}
\end{equation}
It seems that since $H(r)\propto P(r)\propto G(r)^2$, if $r=r_0>0$ is a first-order root of $G(r)$, then it is a second-order root of $H(r)$.
However, the integral within the expression for $P(r)$ exhibits a logarithmic divergence. This divergence effectively reduces the multiplicity of the root by one.
Assuming that near $r=r_0$, $G(r)$ can be expanded as
\begin{equation*}
G(r)= G'(r_0)(r-r_0)+\frac{G''(r_0)}{2}(r-r_0)^2 + \mathcal{O}(r-r_0)^3,
\end{equation*}
substitute it into the solution of $H(r)$, we have
\begin{equation*}
H(r)=\frac{c_1}{r_0^2}(r-r_0)
+\frac{1}{2r_0^3}\Big(\frac{16 \lambda  r_0-8 c_1 (\alpha +2 \lambda )}{\alpha }-\frac{3 c_1 r_0 G''\left(r_0\right)}{G'\left(r_0\right)}\Big)(r-r_0)^2
+ \mathcal{O}(r-r_0)^3.
\end{equation*}
If $c_1\neq 0$, then $r=r_0$ is also first-order root of $H(r)$.
The outcome is different when $r=r_0$ possesses a multiplicity of two in $G(r)$.
With $G(r_0)=G'(r_0)=0$, one can check
\begin{equation*}
H(r)=\frac{c_1}{4 r_0^2}
-\frac{c_1}{6 r_0^3}\Big(\frac{12 \lambda }{\alpha }+\frac{r_0 G^{(3)}\left(r_0\right)}{G''\left(r_0\right)}\Big)(r-r_0)
+\mathcal{O}(r-r_0)^2,
\end{equation*}
and for general $c_1$, $r=r_0$ is not the root of $H(r)$.

Conversely, the situation is different: if $r=r_0$ is a first-order root of $H(r)$, it does not necessarily follow that $G(r_0) = 0$.
If $H(r_0)=0$ and $H'(r_0)\neq 0$, one can check that from \eq{eq:BGH}, apart from the solutions where $G(r_0) = 0$, there also exist a choice 
\begin{equation*}
G(r_0) = -\frac{\alpha  r_0^2 G'\left(r_0\right) H'\left(r_0\right)}{4 \left(\alpha  r_0 H'\left(r_0\right)-2 \lambda \right)}.
\end{equation*}
Such a behavior is exactly what is represented in the solution~(\ref{eq:strangesol}).

\section{\texorpdfstring{An exact four-parameter solution for the case $\xi=0$ and $V\equiv0$}{An exact four-parameter solution for the case xi=0 and V=0}}\label{sec:4sol}

The solution is
\begin{equation}
\begin{aligned}
ds^2=&-\frac{\left(R_s-2 Q\right){}^2 (Q+r-\Delta )^2 }{\left(R_s+2 Q\right){}^2(Q-r+\Delta )^2 }dt^2
+\left(1-\frac{R_s}{r}+\frac{Q^2}{r^2}\right)^{-1}dr^2
+r^2d\Omega^2 \\
=&-\frac{\left(R_s-2 Q\right){}^2 (Q+r-\Delta )^2 }{\left(R_s+2 Q\right){}^2(Q-r+\Delta )^2 }dt^2
+\frac{r^2}{\Delta^2}dr^2
+r^2d\Omega^2,\label{eq:strangesol}
\end{aligned}
\end{equation}
\begin{equation}
b_t(r)=c_1\frac{Q-\Delta}{r}+c_2,
\end{equation}
\begin{equation}
\begin{aligned}
b_r(r)=
\Biggl[
&
\frac{r^2 \left(4 Q^2 \left(-1+\left(c_1{}^2+c_2{}^2\right) \lambda \right)+4 Q \left(R_s+2 c_1 c_2 \lambda  R_s\right)+\left(-1+\left(c_1{}^2+c_2{}^2\right) \lambda \right) R_s^2\right)}{\Delta ^2 \lambda  \left(R_s-2Q\right){}^2}
\\&
+r\Big(
\frac{c_1{}^2 \left(\kappa  \left(R_s^2-4 Q^2\right){}^2-4 \lambda  R_s^2 \left(4 Q^2+R_s^2\right)\right)-8 c_2 c_1 \lambda  Q R_s \left(4 Q^2+3 R_s^2\right)-32 c_2{}^2 \lambda  Q^2 R_s^2}{4 \Delta ^2 \lambda  R_s\left(R_s-2 Q\right){}^2}
\\&
\quad\quad\quad +\frac{c_1{}^2 \left(\kappa  \left(R_s^2-4 Q^2\right){}^2-16 \lambda  Q^2 R_s^2\right)-8 c_2 c_1 \lambda  Q R_s \left(4 Q^2+R_s^2\right)-16 c_2{}^2 \lambda  Q^2 R_s^2}{4 \Delta  \lambda  Q R_s \left(R_s-2 Q\right){}^2}
\Big)
\\&
+\frac{2 Q (\Delta +Q) \left(2 c_2 Q+c_1 R_s\right){}^2}{\Delta ^2 \left(R_s-2 Q\right){}^2}
\Biggl]^{\frac{1}{2}},
\end{aligned}
\end{equation}
where
\begin{equation}
\Delta=\pm\sqrt{r^2-R_s r +Q^2},
\end{equation}
and $R_s$, $Q$, $c_1$ and $c_2$ are free parameters.
The plus and minus signs in $\Delta$ represent two distinct branches of the solution.
If the parameters $c_1$ and $c_2$ are chosen to be zero, a simpler choice of the vector field to support the metric is
\begin{equation*}
b_t(r)=0, \quad b_r(r)=\frac{1}{\sqrt{-\lambda}}\frac{r}{\Delta},
\end{equation*}
hence we need $\lambda<0$.
An interesting aspect of the solution is that if $c_1=0$, then we have $B_\mu B^\mu=-\frac{1}{\lambda}$.
This solution remained unidentified in Sec.~\ref{sec:solve} due to $s(\ell_1+\ell_2)-1 = -\lambda B_\mu B^\mu-1\equiv 0$, and as a consequence \eq{eq:Req} cannot result in $R(r)=r$.
The Kretschmann scalar is
\begin{equation}
K=R_{abcd}R^{abcd}=\frac{40 Q^4+8 Q^2 r \left(3 r-5 R_s\right)+6 r^2 R_s^2-16 Q^3 \Delta}{r^8},
\end{equation} 
showing that the poles of $g_{rr}$ are not singularities.

If $\Delta=0$ has no solutions ($R_s<2Q$), then the metric is a naked singularity.
If $R_s = 2Q$, by first performing an asymptotic expansion of $g_{tt}$ in the limit $r \to \infty$, and subsequently taking the extremal limit $R_s \to 2Q$ for the expansion coefficients, we find that $g_{tt}$ and $1/g_{rr}$ coincide. This specific limiting procedure reveals that the metric reduces precisely to that of an extremal RN black hole, and it is still a naked singularity.

If the equation $\Delta=0$ has two solutions $r_{+}$ and $r_{-}$ ($r_{+} >r_{-}$), the seamless splicing of these two solutions at the throat $r=r_+$ characterizes a traversable Morris–Thorne wormhole.
To see that, denote $\frac{r^2}{\Delta^2}dr^2=dl^2$, we find that $r \approx r_+ + \alpha_1 l^2$ and $\Delta \approx \pm \alpha_2 |l|$ as $r \to r_+$. 
To obtain a globally well-defined manifold, the metric must be analytically continued to the domain where $l < 0$.
However, on a single branch, $g_{tt}$ is a smooth function of $|l|$ rather than $l$. Nevertheless, by splicing the two branches, assigning the positive and negative signs to opposite sides of the throat, one achieves a smooth matching that constitutes a complete manifold.
Therefore, the integration of both branches is essential to form a global manifold, which effectively describes the spacetime of a traversable wormhole.


\bibliography{refs}

\end{document}